\documentclass[journal,transmag]{IEEEtran}
\usepackage[utf8]{inputenc}
\usepackage{amsmath}
\usepackage{amsfonts}
\usepackage{amssymb}
\usepackage[normalem]{ulem}
\usepackage{enumerate}
\usepackage{booktabs}
\usepackage[svgnames]{xcolor}
\usepackage{mathtools}
\usepackage{marvosym}
\usepackage{comment}
\usepackage{siunitx}

\mathtoolsset{centercolon}
\usepackage[colorlinks=true, linkcolor=blue, urlcolor=blue, citecolor=blue]{hyperref}
\usepackage{cleveref}
\crefname{figure}{Fig.}{Figs.} 
\Crefname{figure}{Fig.}{Figs.}

\newcommand{\overbar}[1]{\mkern 1.5mu\overline{\mkern-1.5mu#1\mkern-1.5mu}\mkern 1.5mu}
\newcommand{\real}{\mathbb{R}}
\newcommand{\nat}{\mathbb{N}}

\DeclarePairedDelimiter\floor{\lfloor}{\rfloor}

\usepackage{cite}

\ifCLASSINFOpdf
  \usepackage[pdftex]{graphicx}
  \DeclareGraphicsExtensions{.pdf,.jpeg,.png}
\else
\fi

\usepackage{url}

\begin{document}

\title{Self-Validated Learning for Particle Separation: A Correctness-Based Self-Training Framework Without Human Labels}
\author{\IEEEauthorblockN{Philipp D. L\"osel, Aleese Barron,
Yulai Zhang, Matthias Fabian, Benjamin Young,}
\IEEEauthorblockN{Nicolas Francois, and Andrew M. Kingston}

\thanks{
This work was supported by the Australian Research Council via the ARC Training Centre for Multiscale 3D Imaging, Modelling, and Manufacturing (M3D Innovation, project IC 180100008) and was undertaken with the assistance of resources from the National Computational Infrastructure (NCI Australia), an NCRIS enabled capability supported by the Australian Government. Additional support was provided by AMD under the Heterogeneous Accelerated Compute Clusters (HACC) program. We thank Philipp Gerstner for editing the manuscript and providing helpful comments. ({\it Corresponding author: Philipp D. L\"osel)}

Philipp D. L\"osel, Aleese Barron, Yulai Zhang, Benjamin Young, Nicolas Francois, and Andrew M. Kingston are with the Department of Materials Physics, Research School of Physics, The Australian National University; 60 Mills Rd, Canberra, ACT, 2601, Australia.

Philipp D. L\"osel and Matthias Fabian are with the Biomedisa Development Team, Canberra, ACT, 2601, Australia (email: philipp.david.loesel@gmail.com).}}

\IEEEtitleabstractindextext{%
\begin{abstract}
\textit{Abstract}—Non-destructive 3D imaging of large multi-particulate samples is essential for quantifying particle-level properties, such as size, shape, and spatial distribution, across applications in mining, materials science, and geology. However, accurate instance segmentation of particles in tomographic data remains challenging due to high morphological variability and frequent particle contact, which limit the effectiveness of classical methods like watershed algorithms. While supervised deep learning approaches offer improved performance, they rely on extensive annotated datasets that are labor-intensive, error-prone, and difficult to scale. In this work, we propose self-validated learning, a novel self-training framework for particle instance segmentation that eliminates the need for manual annotations. Our method leverages implicit boundary detection and iteratively refines the training set by identifying particles that can be consistently matched across reshuffled scans of the same sample. This self-validation mechanism mitigates the impact of noisy pseudo-labels, enabling robust learning from unlabeled data. After just three iterations, our approach accurately segments over 97\% of the total particle volume and identifies more than 54,000 individual particles in tomographic scans of quartz fragments. Importantly, the framework also enables fully autonomous model evaluation without the need for ground truth annotations, as confirmed through comparisons with state-of-the-art instance segmentation techniques. The method is integrated into the Biomedisa image analysis platform (\url{https://github.com/biomedisa/biomedisa/}).
\end{abstract}

\begin{IEEEkeywords}
Self-Validated Learning, Self-Training, Particle Separation, Instance Segmentation, Self-Consistency Evaluation, Self-Learning System, Without Manual Annotations.
\end{IEEEkeywords}}

\maketitle

\IEEEdisplaynontitleabstractindextext

\IEEEpeerreviewmaketitle

\section{Introduction}

\IEEEPARstart{X}{-ray} micro-computed tomography (micro-CT) is a powerful tool for three-dimensional (3D) particle characterization, capable of imaging thousands of particles in a single scan. Accurate segmentation is essential for various scientific and industrial applications, including materials science, geology, mining, and pharmaceuticals  \cite{ulusoy_review_2023, santamarina_soil_2015, tang_deep_2022}.

Precise segmentation enables detailed morphological analysis, particle interaction tracking, and quantitative measurements critical for understanding material properties, biological processes, and fluid behaviors. Meaningful statistical analysis of local geometrical, mechanical, or biological properties depends critically on this accuracy.

Classical segmentation methods, such as the widely used watershed algorithm \cite{sheppard_techniques_2014, wang_3d_2016, WANG2015185, rettenberger_uncertainty-aware_2024, guo_method_2022, zhang_improved_2024}, often struggle with heterogeneous datasets, particularly when dealing with irregular shapes and varying object sizes. Recent advances in deep learning have significantly improved instance segmentation performance for both particle \cite{rettenberger_uncertainty-aware_2024,ParticleSeg3D} and cell segmentation tasks \cite{2021_maskrcnn_cells, archit_segment_2025}. However, challenges persist when objects touch or blend together, frequently resulting in under-segmentation (multiple particles sharing one label) or over-segmentation (a single particle split into several labels). Moreover, generating high-quality training data or performing manual corrections remains labor-intensive and impractical at large scales.

To mitigate these challenges, data augmentation techniques, such as simulating domain specific processes \cite{WANG2025100072} have been employed to artificially expand training datasets. While effective to some extent, these methods still depend on manually labeled data.

Adapting pre-trained models to downstream tasks has been a core strategy in machine learning since its early days \cite{Thrun-1995-16236}. This paradigm has gained renewed prominence with the rise of large-scale foundation models, which are trained on diverse, extensive datasets and designed to generalize across a wide range of tasks. In computer vision, for example, the Segment Anything Model (SAM) \cite{kirillov_segment_2023} was initially developed for prompt-based interactive segmentation but also achieves high accuracy in fully automatic segmentation on real-world images.

Additionally, pre-trained models such as Mask R-CNN \cite{he_mask_2017} (via Detectron2 \cite{wu2019detectron2}) and YOLOv8 \cite{yolov8_ultralytics}, originally trained on 2D natural images, can also be used for fully automated instance segmentation. In 3D biomedical imaging, the BioImage Model Zoo \cite{ouyang_bioimage_2022} offers a broad selection of pre-trained models, deployable through tools like DeepImageJ \cite{gomez-de-mariscal_deepimagej_2021} or ilastik \cite{berg_ilastik_2019}. For 3D particle segmentation, ready-to-use models have also been published for diverse materials, requiring no additional training \cite{ParticleSeg3D}.

However, pre-trained models are usually limited to their training domain and their performance often significantly deteriorates when applied to data outside their original training distribution. In such cases, it requires task-specific training (from scratch) or fine-tuning the model on a domain-specific dataset. Unsurprisingly, domain-adapted models consistently outperform generic ones in these settings \cite{ma_segment_2024, archit_segment_2025}. Yet, this performance comes at the cost of substantial manual annotation and domain expertise, efforts that are labor-intensive, time-consuming, and difficult to scale, ultimately limiting full automation.

While data augmentation and transfer learning improve model performance in data-scarce settings, both rely on manually labeled annotations, limiting their scalability. In contrast, self-supervised learning (SSL) enables models to learn directly from raw data by deriving supervisory signals automatically, without the need for human-labeled annotations \cite{SurveySSL}.

SSL typically employs pretext tasks such as masked autoencoding \cite{He_2022_CVPR}, rotation prediction \cite{gidaris2018}, and image transformations \cite{noroozi2017}, or adopts contrastive learning strategies \cite{chaitanya2020contrastivelearning} to learn meaningful data representations. While these models are trained using supervised objectives, the supervisory signals are automatically derived from the data itself, without human annotations. This makes SSL particularly valuable in scenarios where labeled data is scarce or unavailable. As a result, SSL is commonly used for pretraining, with models later fine-tuned on specific downstream tasks.

However, building fully autonomous self-learning systems requires moving beyond generic pretext tasks by incorporating task-specific feedback directly into the training process. Self-training (ST), a widely used semi-supervised approach, typically starts with a small set of labeled examples and refines the model by iteratively adding its own predictions on unlabeled data, so-called pseudo-labels \cite{Lee2013PseudoLabelT}, that meet a predefined confidence threshold or validation criterion \cite{AMINI2025128904}. Moreover, advanced approaches combine SSL with semi-supervised learning to further enhance performance \cite{zhai_s4l_2019}. However, self-training still relies on initial ground truth labels, which must be re-established whenever the data distribution shifts. Moreover, pseudo-labels are inherently uncertain, as there is typically no internal or external mechanism to verify their correctness. This lack of validation can cause the model to overfit to noisy labels, risking performance degradation or even collapse \cite{Chapelle2009} (see also our Section~\ref{sec:st}).

\begin{figure}[!t]
\centering
\includegraphics[width=2.5in]{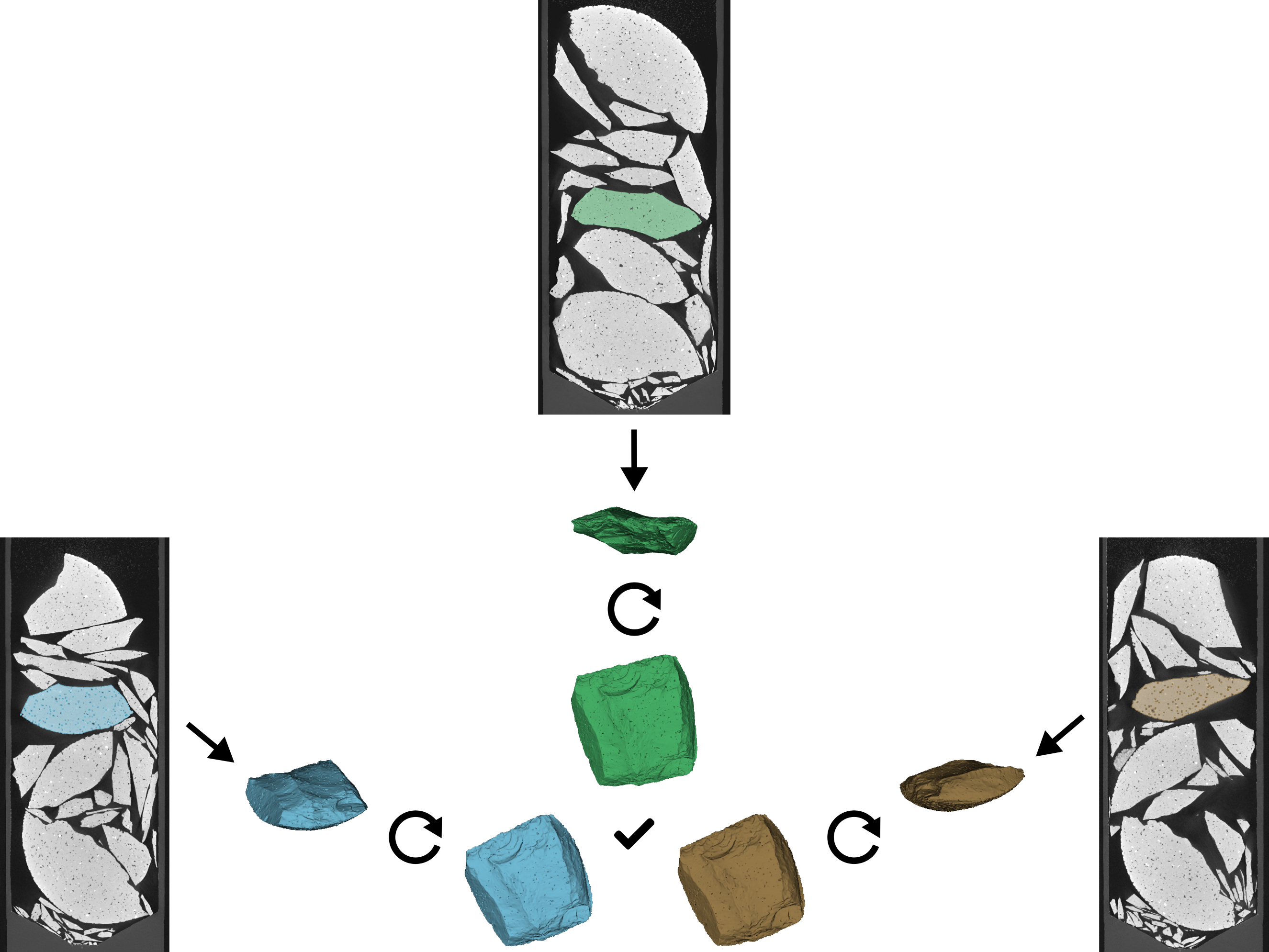}
\caption{Consistent particle identification across reshuffled scans.
The same particle is identified across three independently reshuffled and rescanned volumes of large particles. The 3D segmentations from the first, second, and third scans are shown in light blue, brown, and green, respectively. After rotational alignment, the particle instances are matched, confirming consistency across scans.}
\label{fig:matching}
\end{figure}

\textit{Proposed Method:}
To overcome the limitations of conventional self-training, particularly the accumulation of errors from unverified pseudo-labels, we propose Self-Validated Learning (SVL), a fully autonomous framework that incrementally expands the training dataset and improves model performance without any human annotations. Unlike standard approaches that rely on confidence thresholds or regularization, SVL uses a correctness-based selection strategy: only predictions that are validated through cross-scan instance matching are retained for further training (Fig.~\ref{fig:matching}). This internal validation mechanism ensures high label reliability by confirming consistency across independent observations, without the need for external supervision or ground truth.

A central feature of our approach is implicit boundary detection, which enables training directly from partially labeled data by focusing on confidently segmented regions while naturally excluding unlabeled areas in each iteration (Fig.~\ref{fig:segmentation}).

Our consistency-based evaluation is model-agnostic and supports validation of any segmentation method. It also facilitates method comparison, hyperparameter tuning, and quality assessment under varying imaging conditions, making it a flexible tool for both model development and imaging pipeline optimization.

Our self-matching strategy enables validation of both initial and predicted instance segmentations throughout training. We demonstrate its effectiveness on a large-scale rock particle segmentation task, successfully identifying over 54,000 particles and capturing more than 97\,\% of the total particle volume across three independently reshuffled scans of crushed quartz sphere fragments. All results were achieved entirely without manual annotations. The method is available via Biomedisa \cite{losel_introducing_2020}.

\section{Related Work}

In this work, we build upon the state-of-the-art in instance segmentation and self-training. This section reviews the most relevant advancements in both areas that inform and motivate our approach.

\subsection{Instance Segmentation}
Instance segmentation in particle and cellular datasets often involves explicit boundary detection followed by watershed transforms or connected component analysis. Boundaries can be identified using classical edge detectors like Canny \cite{canny_computational_1986}, machine learning approaches such as Random Forests (e.g., Trainable Weka Segmentation \cite{arganda-carreras_trainable_2017} in Fiji \cite{schindelin_fiji_2012}), or deep learning architectures like U-Net \cite{navab_u-net_2015}. Tools such as PlantSeg \cite{wolny_accurate_2020} and ParticleSeg3D \cite{ParticleSeg3D} use U-Nets to explicitly predict boundaries and separate touching instances.

More advanced methods like DeepWatershed \cite{bai_deep_2017} integrate multiple networks: one predicts unit vectors pointing away from object boundaries, and another reconstructs a distance transform to enable more accurate instance separation. Similarly, Cellpose \cite{stringer_cellpose_2021} predicts a spatial flow field, a vector at each pixel that points from the pixel toward the center of the object it belongs to.

The Segment Anything Model (SAM) employs a transformer-based architecture for prompt-driven segmentation of natural images and has inspired domain-specific adaptations such as MedSAM~\cite{ma_segment_2024} and $\mu$SAM~\cite{archit_segment_2025}. However, these models still require user input (e.g., prompts or bounding boxes) and are computationally demanding, limiting their scalability for fully automated, high-throughput tasks.

Mask R-CNN \cite{he_mask_2017} remains a robust and flexible model for fully automated 2D instance segmentation, especially with fine-tuning from pretrained models on datasets like COCO \cite{COCO} or ImageNet \cite{ImageNet}. YOLOv8 \cite{yolov8_ultralytics} adds segmentation capabilities to the YOLO object detection framework \cite{redmon_you_2015}, but its focus remains on fast object detection rather than precise instance-level separation.

While most of these models offer strong performance on natural image benchmarks, their applicability to 3D volumetric data is limited. Scaling them to domains like materials science or geoscience presents challenges in both computational cost and label availability. Moreover, they lack integrated mechanisms for automated quality control or validation, hindering domain adaptation without manual intervention.

\subsection{Particle Separation}
Particle segmentation has been extensively studied, particularly in geoscience and materials science, with some methods addressing the problem from a purely geometrical perspective \cite{hales_formal_2017}. Existing approaches span classical image processing, machine learning, and deep learning techniques, often in combination. For example, Andrew et al. \cite{andrew_quantified_2018} compared multi-Otsu thresholding, watershed transforms, and a Random Forest-based method implemented in ZEISS Zen Intellesis for segmenting synthetic rock images from sandstone and shale. 

Despite these advances, watershed-based methods remain dominant \cite{sheppard_techniques_2014, wang_3d_2016, WANG2015185, rettenberger_uncertainty-aware_2024}, with enhancements such as adaptive thresholding \cite{guo_method_2022} and post-processing strategies to correct over-segmentation \cite{zhang_improved_2024, kunning2023}.

Deep learning has demonstrated strong performance in both 2D \cite{liu_ore_2020} and 3D \cite{FURAT2024108520} segmentation tasks, but most approaches still rely on manually labeled training data. For example, ParticleSeg3D \cite{ParticleSeg3D} employs nnU-Net \cite{isensee_nnu-net_2021} to automatically segment 3D rock particles. While it is trained on diverse particle types, the method lacks mechanisms for self-adaptation or self-validation on previously unseen, unlabeled datasets, leading to reduced performance when applied outside its training domain.

Overall, traditional approaches struggle with densely packed or morphologically diverse particles, and even deep learning methods offer limited generalizability without labeled data. The absence of built-in validation mechanisms restricts their scalability and reliability. These limitations underscore the need for approaches, like our proposed method, that eliminate manual annotation and introduce internal self-validation, enabling robust, scalable, and fully autonomous instance segmentation of complex particle systems.

\subsection{Self-Training}
Self-training is one of the earliest and most widely used approaches in semi-supervised learning, designed to leverage large amounts of unlabeled data when labeled examples are scarce~\cite{AMINI2025128904}. Originating in the work of Yarowsky (1995) and Nigam et al. (2000)~\cite{yarowsky-1995-unsupervised, nigam_text_2000}, it has since become a foundational technique across domains such as computer vision~\cite{Xie_2020_CVPR, Zoph2020}, natural language processing~\cite{nlp2020}, speech recognition~\cite{st_speech_recognition}, and medical imaging~\cite{bai2017, CHAITANYA2023102792}, where manual annotation is expensive or impractical.

At its core, self-training iteratively expands the labeled dataset by assigning pseudo-labels to unlabeled data. A model trained on labeled samples predicts labels for unlabeled instances, and predictions exceeding a confidence threshold are added to the training set. Typically, the confidence of a pseudo-label is estimated by the probability that the model assigns to its most likely class, which is the predicted class with the highest output score. More sophisticated approaches estimate uncertainty using methods like Monte Carlo dropout~\cite{Uncertainty-aware_self-training2020} or anatomical deviation~\cite{ADIGAV2024103011}. While this can gradually improve performance, it risks accumulating errors: incorrect high-confidence labels can lead to overfitting to noise, performance degradation, or stagnation, especially under domain shift or class imbalance~\cite{Chapelle2009}.

To address these challenges, advanced variants such as Noisy Student Training~\cite{Xie_2020_CVPR} and FixMatch~\cite{FixMatch} combine strong data augmentation with consistency regularization. The core assumption of consistency regularization is that different augmented versions of the same image should yield the same prediction. In FixMatch, for instance, a confident prediction on a weakly augmented image is used to supervise its strongly augmented counterpart. FlexMatch~\cite{FlexMatch} further enhances this approach by dynamically adjusting the confidence threshold for pseudo-labeling.

While conceptually related to our method, these approaches rely on synthetic transformations, such as rotation, flipping, or distortion, that preserve the image structure but do not alter physical relationships between particles. In contrast, we apply physical reshuffling, creating new contact configurations that enable real-world consistency checks.

Although consistency regularization has become increasingly popular, studies like Cascante-Bonilla et al.~\cite{cascantebonilla2020curriculum} demonstrate that classical pseudo-labeling remains competitive.

Despite their promise, self-training methods typically require a small manually labeled seed set, which must be reannotated if the data distribution shifts. Furthermore, they rely on heuristics to estimate label correctness and lack a principled validation mechanism.

\section{Method}
Here, we introduce the concept of Self-Validated Learning (SVL), a correctness-driven self-training framework. SVL addresses key limitations of conventional approaches by verifying pseudo-labels through physically grounded, cross-scan instance matching. Instead of relying on prediction confidence or heuristic constraints, SVL validates a segmentation by ensuring consistency across independently reshuffled scans of the same sample. Only predictions that consistently match across these physically transformed observations are retained, providing a level of correctness not found in traditional self-training methods.

A segmented particle is accepted only if it can be reliably identified and matched in at least one additional reshuffled scan of the same sample (see Fig.~\ref{fig:matching}). These additional micro-CT scans of the sample are undertaken after physically extracting, shuffling, and repacking the particles. These reshuffled scans yield multiple independent views of the same particle population for robust validation.

SVL follows an iterative process that progressively expands the training set with self-validated labels. It begins by training a model to predict particle boundaries. From these predictions, individual particles are segmented by subtracting the predicted boundaries from a binary mask of all particles (particle mask) and labeling the connected components.

Matching proceeds in stages of increasing granularity, first comparing particle volume, then surface-to-centroid distance distributions, and finally mask overlap after aligning candidate pairs to a common orientation. A prediction is accepted if the aligned overlap exceeds a predefined threshold. To further refine a segmentation, majority voting across matched instances is applied. Only these validated particles are added to the training set, while uncertain or unmatched regions are excluded.

In the following subsections, we detail each component of the SVL framework:
\begin{itemize}
\item Section~\ref{sec:segmentation}: Introduces the mathematical and voxel-based representation of micro-CT data and segmentation.
\item Section~\ref{sec:implicit}: Introduces implicit boundary detection, which handles class imbalance and excludes unlabeled regions without the need for an ignore mask.
\item Section~\ref{sec:matching}: Describes our cross-scan matching strategy.
\item Section~\ref{sec:correction}: Explains how majority voting improves segmentation accuracy.
\item Section~\ref{sec:svl_algorithm}: Presents the complete SVL pipeline.
\end{itemize}

\begin{figure*}[!t]
\centering
\includegraphics[width=0.95\textwidth]{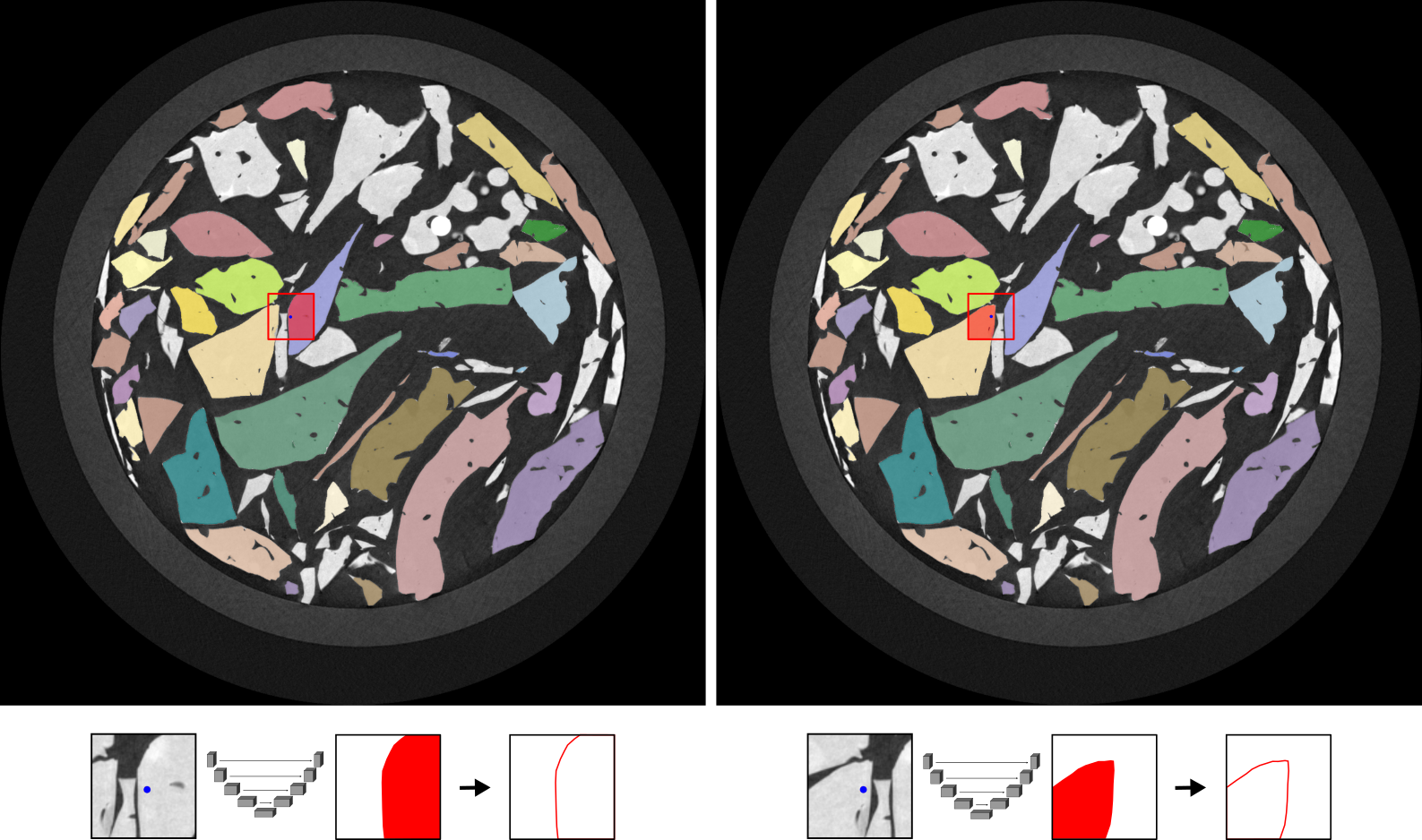}
\caption{Implicit boundary detection for particle segmentation.
A 3D U-Net is trained on micro-CT image data with dynamically generated binary label patches. The figure shows a cross-section of crushed medium-sized rock particles, where matched particles are highlighted (colored) and unmatched particles remain unlabeled. Implicit boundary detection is trained exclusively on patches centered on matched particles.
The red rectangle with a blue center point illustrates a training patch, where segmentation is defined relative to the patch center. Depending on its position, the same particle (e.g., the yellow one on the left) may be treated as background (left) or foreground (right). Note that the center point is for illustration only and is not part of the network input, which consists solely of raw tomographic data. Boundaries are derived from the resulting binary segmentation.}
\label{fig:segmentation}
\end{figure*}

\subsection{Voxel-Based Image Representation and Segmentation}\label{sec:segmentation}
The reconstructed image of a micro-CT scan can be represented as
\begin{align*}
u\colon \Omega \to \mathcal M \text{ with } \Omega\subset\real^3 ,
\end{align*}
where the domaine $\Omega$ denotes the 3D spatial volume and the codomain $\mathcal M\subset \real$ specifies the image intensity values. For 16-bit grayscale data, the intensity range is $\mathcal M=\{0,\ldots,2^{16}-1\}$.
The domain $\Omega$ is partitioned into equally sized cubes,
\begin{align*}\overbar\Omega=\bigcup_{\substack{1\leq i\leq l\\1\leq j\leq m\\1\leq k\leq n}} \overbar\Omega_{ijk},
\end{align*}
with disjoint subsets $\Omega_{ijk} \cap \Omega_{i^\prime j^\prime k^\prime} = \varnothing$ for $(i,j,k)\neq(i^\prime,j^\prime,k^\prime)$. Each of these $l\times m\times n$ cubes, called \textit{voxels}, is assigned a single constant intensity value,
\begin{align*}
\left.u\right|_{\Omega_{ijk}}=\text{const} \in \mathcal M.\end{align*}
A \textit{segmentation} of $u$ can then be described by a mapping
\begin{align*}
s\colon \Omega \to \{0,1,2,\ldots\},
\end{align*}
where each voxel $\Omega_{ijk}$ is assigned a \textit{label}. 
For a binary segmentation mask $m$ distinguishing two regions (e.g., foreground and background), this reduces to
\begin{align*}
  \left.m\right|_{\Omega_{ijk}} =
  \begin{cases}
    1, & \text{if }\Omega_{ijk}\text{ belongs to the foreground}, \\
    0, & \text{if }\Omega_{ijk}\text{ otherwise}.
  \end{cases}
\end{align*}
Accordingly, an individual particle can be represented as a binary mapping 
\begin{align*}
p\colon\Omega \to \{0,1\}.
\end{align*}

\subsection{Implicit Boundary Detection and Separation}\label{sec:implicit}
For implicit boundary detection, label patches of confidently segmented particles are dynamically converted into binary masks during training. The region containing the patch center is treated as foreground, while the rest is considered background (\cref{fig:segmentation}). Only patches whose center lies within a labeled region are used.

A patch \[T=\{ \Omega_{ijk} \ |\ i_0\le i< i_1, j_0\le j< j_1, k_0\le k< k_1 \}\] is a subvolume of size $(i_1-i_0)\times (j_1-j_0)\times (k_1-k_0)$ voxels within the domain $\overbar\Omega$. The label at the patch center is given by
\begin{equation*}
l_T=s\left(\floor*{\frac{i_1-i_0}{2}},\floor*{\frac{j_1-j_0}{2}},\floor*{\frac{k_1-k_0}{2}}\right),
\end{equation*}
where $s$ denotes the segmentation, with $s=0$ indicating background or unlabeled particles. This label is used to define a binary segmentation mask:
\begin{align*}
  m_T(\mathbf{x}; s) =
  \begin{cases}
    1, & \text{if }\mathbf{x}\in T\text{ and }s(\mathbf{x})=l_T, \\
    0, & \text{otherwise}.
  \end{cases}
\end{align*}
Training is performed using standard supervised learning on all patches with $l_T>0$, using the corresponding binary masks, $m_T$. 

During inference, only patches centered within an unlabeled particle region (see Fig.~\ref{fig:63microns_zoom}) are used for boundary prediction. To select these regions, a \textit{positive mask} of unlabeled particles is defined as
\begin{equation*}
\widehat{m}(\mathbf{x}) = m(\mathbf{x}) - \left(\frac{s(\mathbf{x})}{\max(1,s(\mathbf{x}))}\right),
\end{equation*}
where $m$ denotes the mask of all particles. The binary boundary map, $b(\mathbf{x})$, of unlabeled particles over the entire domain $\Omega$ is then constructed by aggregating the gradients of all patch-level masks $m_T(\mathbf{x};\widehat{m})$ with $l_T=1$:
\begin{align}\label{eq:aggregating}
  b(\mathbf{x}) =
  \begin{cases}
    1, & \text{if }\sum_T |\nabla m_T(\mathbf{x};\widehat{m})|>0, \\
    0, & \text{otherwise}.
  \end{cases}
\end{align}
Here, $\nabla m_T(\mathbf{x})$ denotes the discrete gradient of the binary mask $m_T$, which is nonzero only at object boundaries. The predicted boundaries are then subtracted from the overall particle mask, and the remaining connected components are assigned unique labels. Finally, the removed boundary regions are filled using the nearest label according to Euclidean distance (see Appendix~\ref{sec:separation}).

\subsection{Particle Matching}\label{sec:matching}
To match identical particles across reshuffled volumes, we use an hierarchical approach based on increasingly discriminative metrics: particle volume, surface-to-centroid distance histograms (Eq.~\ref{eq:histogram}), and Dice score (Eq.~\ref{eq:dice}). To reduce computational complexity, each particle is first filtered by volume, considering only candidates within $\pm10\,\%$ of its size (Step 1). Among these, the best match is selected by minimizing the mean squared error between surface-to-centroid distance histograms, a well-established translation- and rotation-invariant comparison metric \cite{ma_netra_1999} (Step 2, Fig. 3). Finally, we compute the optimal rotation that maximizes the Dice score (Step 3). Two particles are considered a match if the resulting Dice score exceeds a defined threshold (e.g., 0.9).

For a particle, $p\colon\Omega \to \{0,1\}$, the set of voxels belonging to the particle is:
\begin{align*}
P= \{ \Omega_{xyz}\ \text{with}\ x,y,z\in \nat \ |\ p(x,y,z)=1\}.
\end{align*}
The centroid, $\mathbf{c}\in\real^3$, is given by:
\begin{align*}
\mathbf{c}=\frac{1}{|P|} \sum_{\Omega_{xyz}\in P} (x,y,z).
\end{align*}
Let $\mathcal S=\{\mathbf{s}_i\}_{i=1}^N$ be the set of $N$ particle surface points in $\real^3$.
The Euclidean distance from the centroid, $\mathbf{c}$, to each surface point, $\mathbf{s}_i$, is found as:
\begin{align*}
d_i=\|\mathbf{s}_i-\mathbf{c}\|=\sqrt{(x_i-x_\mathbf{c})^2 + (y_i-y_\mathbf{c})^2 + (z_i-z_\mathbf{c})^2}
\end{align*}
for all $i=1,\ldots,N$.
The surface-to-centroid distance histogram is given by:
\begin{align*}
H_p=(h_1,h_2,\ldots,h_M),
\end{align*}
where we choose the number of histogram bins as:
\begin{align*}
M=\left\lfloor \frac{d_{\max}-d_{\min}}{2} \right\rfloor,
\end{align*}
and
\begin{align*}
h_j=\sum_{i=1}^N \mathbf{1}_{b_j\leq d_i< b_{j+1}}
\end{align*}
is the number of distances, $d_i$, falling within each bin, $B_j=[b_j,b_{j+1})$, where $\mathbf{1}$ is the indicator function:
\begin{align*}
  \mathbf{1}_{b_j\leq d_i< b_{j+1}} =
  \begin{cases}
    1, & \text{if }b_j\leq d_i< b_{j+1}, \\
    0, & \text{otherwise}.
  \end{cases}
\end{align*}

For each particle, $p$, in a particle pack, we identify a corresponding particle, $q$, in a reshuffled pack (containing the same particles) by selecting candidates with volumes within $\pm10\,\%$ of the volume of $p$ and minimizing the difference between their surface-to-centroid distance histograms:
\begin{align}\label{eq:histogram}
q^\ast=\arg\min_q\|H_p-H_q\|_2.
\end{align}

To compare shapes, both $p$ and the candidate $q^\ast$ are centered at the origin via affine transformation. After filling internal inclusions, we compute Dice scores (Eq.~\ref{eq:dice}) between $p$ and all rotated versions of $q^\ast$ around its centroid.

The voxel set of $q^\ast$ after applying a rotation, $R_{\boldsymbol{\phi}} \colon \real^3 \to \real^3$, with rotation angles $\boldsymbol{\phi}=(\alpha,\beta,\gamma)$ about the $z, y, x$ axes, respectively, is given by:
\begin{align*}
Q^\ast_{\boldsymbol{\phi}}= \{\Omega_{x^\prime y^\prime z^\prime},(x^\prime,y^\prime ,z^\prime)=R_{\boldsymbol{\phi}}(x,y,z)\ |\ q^\ast(x,y,z)=1\}.
\end{align*}

The Dice score between voxel sets $P$ and $Q^\ast_{\boldsymbol{\phi}}$ measures the overlap:
\begin{align}\label{eq:dice}
\mathrm{Dice}(P,Q^\ast_{\boldsymbol{\phi}})=\frac{2|P\cap Q^\ast_{\boldsymbol{\phi}}|}{|P| + |Q^\ast_{\boldsymbol{\phi}}|} \in [0,1],
\end{align}
and the optimal rotation maximizing this score is given by:
\begin{align}
\label{eqn:rotation}
\boldsymbol{\phi}^\ast=\arg\max_{\boldsymbol{\phi}}\mathrm{Dice}(P, Q^\ast_{\boldsymbol{\phi}}).
\end{align}

In the ideal case where $\mathrm{Dice}(P,  Q^\ast_{\boldsymbol{\phi}^\ast})=1$, the particles are identical after rotation:
\begin{align*}
p(\mathbf{x})=q^\ast\left(R_{\boldsymbol{\phi}^\ast}(\mathbf{x})\right)\ \forall\ \mathbf{x}\in\Omega\cap\nat^3.
\end{align*}
A match is recorded if the maximum Dice score after rotation (RotDice) exceeds a given threshold. To minimize computational cost, all rotations and Dice score calculations are executed on GPUs using CUDA and PyCUDA \cite{klockner_pycuda_2012}. Matches from previous iterations are reused to eliminate redundant computations. For particles exceeding 10,000 voxels, both $p$ and $q^\ast$ are uniformly resized such that $p$ contains exactly 10,000 voxels before evaluating RotDice. The optimal rotation obtained at this reduced resolution is then refined at the original resolution, but only within a narrowed rotation interval, further reducing the overall computational effort.

\begin{figure}[!t]
\centering
\includegraphics[width=2.5in]{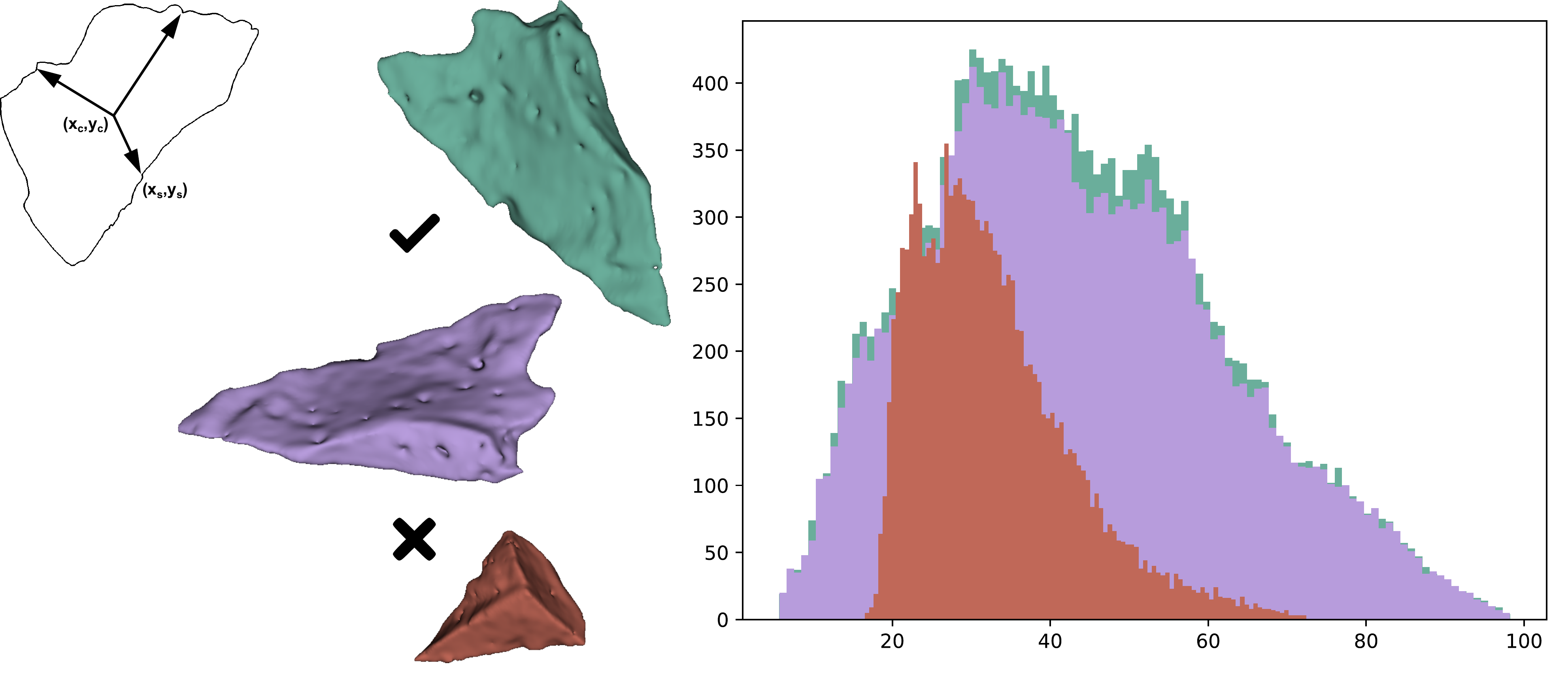}
\caption{Surface distance histograms for different particles. a Illustration of two-dimensional centroid to boundary distances. b Surface renderings of two matching particles (green and violet) and one not matching particle (red) of similar volume (98.3$\%$ of violet’s volume). The green and red particles were extracted after shuffling and rescanning the particle pack containing the violet particle. c Distances measured from the centroid to the surface of each particle from (b) and binned into histograms.}
\label{histogram}
\end{figure}

\subsection{Particle Correction}\label{sec:correction}
After identifying the same particle across multiple scans, the segmentation can be refined by removing false positive voxels and adding false negative voxels using the combined information from all matched instances.

\subsubsection{Three and more matched particles:}
When a particle is matched across $n\geq 3$ scans, correction can be performed via a majority vote. For $i=1,\ldots,n$, let $p_i$ denote a particle from one scan, and let $p_j$, with $j\neq i$, be the corresponding match from another scan. Each pair $(p_i,p_j)$ is aligned using the optimal rotation, $R_{ij}:=R^{p_ip_j}_{\boldsymbol{\phi}^\ast}$, obtained with Eq.~\ref{eqn:rotation}.
The corrected particle, $\hat{p}_i$, is then defined as:
\begin{align*}
  \hat{p}_i(\mathbf{x}) =
  \begin{cases}
    1, & \text{if }\sum_{j\neq i}p_j(R_{ij}(\mathbf{x}))\geq\frac{1}{2}n, \\
    0, & \text{otherwise}.
  \end{cases}
\end{align*}
Note that when $n$ is even, a majority threshold of $\frac{1}{2}n$ may allow ties. To enforce a stricter consensus, the threshold can be raised to $\sum_{j\neq i}p_j(R_{ij}(\mathbf{x}))>\frac{1}{2}n$, requiring more than half of the results to agree.

To prevent altering already corrected particles, a lock mask is maintained for each scan and updated after each correction:
\[L_i \leftarrow L_i \lor \hat{p}_i,\]
starting from $L_i=0$.

\subsubsection{Two matched particles:}
For particles $p_1$ and $p_2$ identified in only two of three or more scans, with corresponding binary segmentation masks $m_1$ and $m_2$ of all particles, false positives in $p_i$ that overlap with background regions in the other scan $(m_j=0)$ are removed. Conversely, false negatives are recovered from the matched particle, provided the corresponding voxel is not already locked. The corrected particles are given by:

\begin{align*}
\hat{p}_1&=m_2(R_{12}(\mathbf{x}))p_1 + p_2(R_{12}(\mathbf{x}))m_1(1-p_1)(1-L_1), \\
\hat{p}_2&=m_1(R^{-1}_{12}(\mathbf{x}))p_2 + p_1(R^{-1}_{12}(\mathbf{x}))m_2(1-p_2)(1-L_2). \\
\end{align*}

\subsection{Self-Validated Learning via Reshuffling and Matching}\label{sec:svl_algorithm}
In each iteration, the unmatched particles from the previous step are segmented and incorporated into the training set once validated, i.e., when the segmented particle is successfully matched across two (or more) micro-CT scans with a matching score exceeding a defined threshold.
Adding more validated particles progressively enriches the training data, which in turn improves segmentation accuracy and enables increasingly reliable statistical and morphological analyses.
The main steps of the method can be summarized as follows:
\begin{enumerate}[1.]
\item \textbf{Detect Boundaries}: Detect particle boundaries using a pretrained model and separate individual particles in reshuffled scans of the same particles.
\item \textbf{Match Particles}: Identify the best-matching pairs based on size and the histogram of boundary distances to the centroid.
\item \textbf{Optimize Rotation}: Determine the rotation that maximizes the Dice score (RotDice).
\item \textbf{Selection Criteria}: Include a particle pair in the training set if its RotDice meets the threshold.
\item \textbf{Correction}:
If at least three scans exist, apply a majority vote: retain or remove areas present in at least $50\%$ of the results.
If only two scans exist, adjust areas based on mutual presence or absence in the particle mask.
\item \textbf{Train Model}: Update the model using correctly identified particles while ignoring unmatched regions.
\end{enumerate}
Correction is skipped for particles already identified in all scans and corrected in a previous iteration. This process repeats until all or a sufficient number of particles are segmented. 

\begin{table*}[!t]
\centering
\caption{\label{tab:comparison_methods} Quantitative comparison of different methods}
\setlength{\tabcolsep}{6pt}
\begin{tabular}{ccccccccccccc}
\toprule
{} &  \multicolumn{3}{c}{Large} & \multicolumn{3}{c}{Medium} & \multicolumn{3}{c}{Small}\\
\midrule
{}   & Particles (\#) & Volume (\%) & Time (min) & Particles (\#) & Volume (\%) & Time (min) & Particles (\#) & Volume (\%) & Time (min) \\ 
SAM ViT-L\cite{kirillov_segment_2023} & 192 & 99.63 & 170 & 2736 & 96.25 & 143 & 27681 & 74.31 & 522 \\
AdvWatershed \cite{zhang_improved_2024} & 179 & 98.69 & 30 & 2850 & 96.24 & 18 & 41826 & 89.48 & 18 \\
SVL@1   & 215 & 99.64 & 2 & 2997 & 98.12 & 5 & 47228 & 95.34 & 21 \\
SVL@2   & 222 & 99.64 & \bf{2}  & 3022 & 98.22 & \bf{3} & 48743 & 96.20 & \bf{14} \\
SVL@2.5 & \bf{230} & \bf{99.67} & 13 & \bf{3057} & \bf{98.38} & 17 & \bf{50942} & \bf{97.08} & 89 \\
\bottomrule
\end{tabular}\\
\raggedright
Summary of matched particles and inference times by method and particle size class. The table reports the total number of matched particles across all three scans, their combined volume as a percentage of the total particle volume, and the average inference time. “@” indicates the iteration step. SVL results were generated at half resolution unless stated otherwise; “@2.5” denotes reapplication of “@2” at full resolution. Bold values indicate the best performance in each column.
\\
\par
\end{table*}

\section{Results and Analysis}
In this section, we present and analyze experiments and results of Self-Validated Learning (SVL) applied to particles of three different size ranges (see Appendix \ref{sec:sampleprep} for details): \textbf{large} (particles with diameter $>$\SI{1}{\milli\meter}, \cref{fig:matching,fig:1mm}), \textbf{medium} (particles with diameter from \SI{1}{\milli\meter} to \SI{250}{\micro\meter}, \cref{fig:segmentation,fig:250microns}), and \textbf{small} (particles  with diameter from \SIrange{250}{63}{\micro\meter}, \cref{fig:63microns}). We compare SVL to an advanced watershed segmentation (Section \ref{sec:svl_results}), a customized SAM-based approach (Section \ref{sec:svl_results}), and standard self-training (Section \ref{sec:st}). Additionally, we evaluate the benefits of implicit boundary detection versus direct boundary prediction (Section \ref{sec:implicit_explicit}).

\subsection{Datasets}
Each particle size group was packed into a separate container and scanned via X-ray micro-computed tomography at voxel resolutions of 5.38, 2.8, and \SI{1.54}{\micro\meter}/voxel for the large, medium, and small particles, respectively. For each group, three independent scans were acquired after physically removing, reshuffling, and reloading the particles. The resulting 3D volumes had average dimensions of $2450\times2450\times5783$ voxels (large), $2342\times2342\times5544$ (medium), and $2730\times2730\times11802$ (small).

\subsection{Reporting and Evaluation of Results}\label{sec:accuracy}
This section explains how segmentation performance is reported throughout the following sections, tables, and figures. To quantitatively evaluate accuracy, we report:
\begin{enumerate}[(a)]
    \item the number of matched particles, and
    \item the matched particle volume as a percentage of the total particle volume (all particles in the packing).
\end{enumerate}
A particle is considered matched if it can be correctly identified in at least one other scan. This means some particles may be matched in only two of the three rescans.

In Figs.~\ref{fig:63microns}–\ref{fig:1mm}, we show particles from the currently displayed scan that were successfully matched in at least one other scan. The first value shown reflects the matched particles in the displayed scan, while the value in parentheses indicates additional particles identified exclusively between the other two scans. Particles in the displayed scan that could not be matched to any other scan are shown in red.

The summary statistics reported in Tables~\ref{tab:comparison_methods}–\ref{tab:explicit_implicit} and subsequent sections include the total number of matched particles and their cumulative volume across all three scans of the same particle pack. These totals represent the combined results of both direct matches (with the scan shown) and complementary matches (between the other two scans). The same procedure applies to both count-based and volume-based analyses.

\begin{table*}[!t]
\centering
\caption{\label{tab:st_vs_svl} Quantitative comparison of Classical Self-Training and Self-Validation Learning}
\setlength{\tabcolsep}{6pt}
\begin{tabular}{ccccc|cccccccc}
\toprule
{} & \multicolumn{2}{c}{Medium} & \multicolumn{2}{c}{Small} & \\
\midrule
{}   & Particles (\#) & Inference (sec) & Particles (\#) & Inference (sec) & Training (min)\\ 
ST@1 \cite{bai2017}  & 2773 & 1634 & 19914 & 4948 & 2156 \\
ST@2 \cite{bai2017}  & 2854 & 1666 & 6 & 5020 & 2186 \\
ST@3 \cite{bai2017}  & 2844 & 1660 & 248 & 5011 & 2214 \\
\midrule
SVL@1  & 2773 & 1634 & 19914 & 4948 & 2156 \\
SVL@2  & 2942 & 315 & 42473 & 2565 & 1598 \\
SVL@3  & \textbf{2981} & \textbf{174} & \textbf{46795} & \textbf{996} & 2074 \\
\bottomrule
\end{tabular}\\
\raggedright
Comparison of particle segmentation progress across iterations using conventional self-training (ST) and self-validated learning (SVL). Both methods were initialized with validated SAM results for large particles (approx.\ 200 particles), simulating a small manually annotated dataset. In ST, all pseudo-labels of unlabeled particles were added to the training set after each iteration. In SVL, only particles correctly matched across rescans were added and subsequently excluded from further prediction updates. Note that in practice, ST cannot be evaluated without an independent validation mechanism, here provided by our SVL method. Bold values indicate the best performance in each column.\\
\par
\end{table*}

\subsection{Benchmark Methods for Instance Segmentation}
To establish a baseline for comparison against Self-Validated Learning (SVL), we evaluated two established instance segmentation methods on all nine micro-CT scans. These scans covered three particle size classes (large, medium, and small) each imaged three times following physical reshuffling. The first method applied the Segment Anything Model (SAM) across all datasets (Appendix~\ref{sec:sam}), while the second employed an advanced watershed algorithm (Appendix~\ref{sec:watershed}).

Using our consistency-based evaluation with a minimum RotDice threshold of 0.9, we matched:
\begin{itemize}
    \item 192 large particles (99.63\,\% of total volume),
    \item 2,736 medium particles (96.25\,\%), and
    \item 27,681 small particles (74.31\,\%)
\end{itemize}
from the SAM results. However, this came at a high computational cost: 170 minutes for large, 143 minutes for medium, and 522 minutes for small particles (Table~\ref{tab:comparison_methods}). In comparison, applying our matching strategy to the output of the advanced watershed method yielded:
\begin{itemize}
    \item 179 large particles (98.69\,\%),
    \item 2,850 medium particles (96.24\,\%), and
    \item 41,826 small particles (89.48\,\%),
\end{itemize}
with significantly reduced inference times: 30 minutes for large particles and 18 minutes each for medium and small.

While SAM excelled in segmenting large particles, its performance declined significantly for smaller ones. This may be improved by extending the tiling strategy used in the $z$-direction to the $x$- and $y$-axes, though at the cost of even longer runtimes. Alternatives such as SAM2 \cite{ravi2024sam2segmentimages}, which promises faster inference, or domain-specific fine-tuning, could further enhance performance. Leveraging video-based propagation strategies, as implemented in SAM2, may also improve the 3D consistency of segmentations when applying 2D models to volumetric data.

In contrast, the advanced watershed method delivered more balanced performance across particle sizes, particularly outperforming SAM for small particles, while being considerably more computationally efficient and requires no GPU acceleration.

However, like all state-of-the-art instance segmentation methods, both approaches lack the ability to validate their results and iteratively improve segmentation performance in dense particle systems. This limitation highlights the need for approaches like SVL, which combine iterative refinement with autonomous validation and scalability.

\subsection{Self-Validated Learning (SVL)}\label{sec:svl_results}
For each particle group (small, medium, and large), we selected the previous segmentation result with the highest number of matched particles to train our first Self-Validated Learning model (SVL@1). Specifically, we used the SAM-based segmentation for the large particle set and the advanced watershed results for the medium and small particle sets.

To reduce computational load, model training (Appendix~\ref{sec:training}) and boundary prediction (Appendix~\ref{sec:inference}) were initially performed on data downsampled by a factor of two along each axis. In the final iteration, the model was applied at full resolution to segment the remaining, more challenging particles. The average dimensions of the downsampled scans were $1225 \times 1225 \times 2892$ voxels for the large particle group, $1171 \times 1171 \times 2772$ for the medium group, and $1365 \times 1365 \times 5901$ for the small group.

The trained SVL@1 model (with "@" denoting the SVL iteration) was then applied to the positive mask (Fig. \ref{fig:63microns_zoom}, Appendix~\ref{sec:separation}) of previously unlabeled particles in each scan. After predicting boundaries, separating touching particles, and identifying consistent instances across all three reshuffled scans, we recovered an additional 23, 147, and 5,402 particles for the large, medium, and small datasets, respectively, augmenting the initial results obtained from SAM and the watershed method (Table~\ref{tab:comparison_methods}).

These newly identified particles were added to the training set, and the model was retrained (SVL@2). Applying this updated model to the remaining unlabeld regions further increased the total number of matched particles to 222, 3,022, and 48,743 for the large, medium, and small groups, respectively.

Finally, we applied the SVL@2 model at full resolution (SVL@2.5), identifying an additional 12, 35, and 2,199 particles. This resulted in total particle volume coverage of 99.67\,\%, 98.38\,\%, and 97.08\,\% for the respective groups. In total, the method successfully segmented over 54,000 particles from the original crushed sphere, significantly exceeding what conventional watershed approaches could achieve, and captured nearly the entire particle volume.

Inference times for SVL@2 decreased slightly due to the shrinking set of unlabeled particles, requiring only 2, 3, and 14 minutes for the large, medium, and small datasets, respectively, using a multi-GPU setup (see Appendix~\ref{sec:inference}). In contrast, full-resolution inference in SVL@2.5 was more computationally demanding despite the reduced unlabeled volume, with runtimes of 13, 17, and 89 minutes, respectively.

\subsubsection{Training on Full-Resolution}
We also tested training on full-resolution data for the small particle class. However, due to host memory limitations, training and inference had to be performed separately on each of the three reshuffled scans. This restricted the model’s ability to learn from cross-scan variability, leading to reduced performance: starting from the SVL@2 result and training a third iteration this way yielded only 50,730 validated particles, compared to 50,942 when SVL@2, trained on all scans at half resolution, was reapplied at full resolution (Table~\ref{tab:comparison_methods}).

\subsubsection{Particle Count vs. Volume Coverage}
Interestingly, while SAM achieved slightly higher volume coverage for medium-sized particles compared to the advanced watershed method (96.25\,\% vs. 96.24\,\%), it identified significantly fewer individual particles. This discrepancy arises from SAM detecting a few large particles that the watershed method missed. Similarly, for large particles, the total segmented volume changed little between SAM and SVL@2.5. However, the number of detected particles increased by nearly 20\,\%, primarily due to the additional identification of very small particles that were previously missed.

\subsubsection{SAM as an Alternative Backbone for SVL}\label{sec:sam_backbone}
Although the Segment Anything Model (SAM) can, in principle, be fine-tuned using Self-Validated Learning (SVL) in combination with an ignore mask (Appendix~\ref{sec:explicit}), it is primarily designed for interactive 2D segmentation and benefits most from its extensive pretraining on diverse datasets. Given the availability of our own verified training data, we opted for a lightweight 3D U-Net variant, which can be trained more efficiently and deployed at scale with significantly faster inference times (Table~\ref{tab:comparison_methods}).

In principle, SAM could be adapted for SVL by fine-tuning its prompt encoder using verified segmentations and randomly sampled point prompts. However, due to SAM’s significant computational overhead, we did not pursue this approach in the current study. Future work may explore whether a dedicated “Segment Any Rocks” model can offer meaningful performance improvements over our efficient implicit method. Notably, at comparable resolution, SAM’s inference was up to $6\times$ slower than our model when using the complete particle mask—averaging 38 minutes for large particle scans, 28 minutes for medium, and 83 minutes for small.

While SAM’s performance could be improved using the same tiling strategy in the $xy$-plane as used along the $z$-axis (Appendix~\ref{sec:sam}), doing so increases the number of tiles per slice from 1 to 9 for small particles, resulting in a ninefold increase in computational cost, further compounding SAM’s already long runtime.

Despite these limitations, SAM may still be useful for initializing the first set of segmented particles to seed our SVL pipeline. However, we demonstrated that the advanced watershed method performs significantly better, particularly for small particles, albeit requiring non–open-source software.

Technically, both implicit boundary detection and the SAM-based approach follow a similar strategy: they predict segmentations (2D instances for SAM, 3D binary volumes for the implicit method), from which boundaries are derived and aggregated into a final 3D result. Since both rely on consistently detected boundaries, they tend to under-segment, i.e., over-merge neighboring particles, especially when thin contact regions are not properly separated. Conversely, they rarely over-segment particles (i.e., split a single particle into multiple labels).

\subsection{Self-Training Without SVL}\label{sec:st}
We compared our proposed Self-Validated Learning (SVL) approach, where only verified segmentations are added to the training set, with conventional self-training (ST), where all pseudo-labels are accepted and updated in each iteration without validation~\cite{bai2017}. Traditional ST assumes access to a small, manually labeled dataset; we simulate this by using the validated SAM results of large particles as initialization for both ST and SVL. This subset contains around 200 particles, an amount still feasible for manual correction or validation.

Both methods were trained for three iterations (Table~\ref{tab:st_vs_svl}). In the first iteration (ST@1/SVL@1), the training procedure was identical for both approaches, as no filtering had been applied yet. This initial model was trained exclusively on the large particle dataset for 15 epochs. The resulting model was then used to predict boundaries for the unlabeled medium and small particle groups.

In subsequent iterations, the strategies diverged. For ST@2 and ST@3, all predicted pseudo-labels were directly added to the training set without any validation. In contrast, SVL@2 and SVL@3 incorporated only correctly matched particles, as identified through our consistency-based validation. Each of these later models was trained for 8 epochs.

In both settings, the model was not re-applied to the large particle group after initialization. Instead, these particles were kept fixed in each iteration, and the model was applied only to the medium and small particles. SVL consistently improved performance across iterations for both groups. In contrast, ST achieved limited improvement for medium particles in the second iteration and failed in the third. For small particles, ST collapsed immediately at ST@2, highlighting its sensitivity to erroneous pseudo-labels.

While we simulate manual initialization, it’s important to note that the validated SAM-based labels would not have been feasible without our self-validation method. In practice, manual annotation or verification is unfeasible for small particles due to their sheer quantity.

SVL also reduces inference time in later iterations by restricting predictions to previously unlabeled regions, for example, inference for small particles dropped from 4,948 seconds in the first iteration to 996 seconds in the third (Table~\ref{tab:st_vs_svl}). In contrast, standard self-training (ST) processes the entire volume in every iteration, resulting in a constant inference time regardless of label progress.

While SVL’s training time increases moderately from 1,598 to 2,074 minutes between the second and third iterations due to the expanding training set, it also introduces computational overhead for matching and filtering. Calculating RotDice scores between two large particle scans takes 5.2 minutes, and for medium particles, 37.3 minutes (SVL@1) or 7.2 minutes (SVL@3) using four NVIDIA V100 GPUs. However, for the small particle scans, the process is significantly longer, 8.8 hours for SVL@1 and 3.4 hours for SVL@3. This additional cost, however, ensures more reliable labels and prevents redundant processing of already-validated regions.

Overall, our results clearly demonstrate the importance of label filtering. SVL substantially outperforms standard ST in both accuracy and robustness, particularly in the challenging case of small-particle segmentation. Without label filtering, ST quickly deteriorates, identifying only 6 particles in the first iteration and recovering only slightly to 248 matched particles thereafter. In contrast, SVL consistently improves across iterations, ultimately identifying 46,795 particles after three iterations (Table~\ref{tab:st_vs_svl}).

\subsection{Implicit vs. Explicit Boundary Detection}\label{sec:implicit_explicit}
We evaluated implicit and explicit boundary detection within our SVL framework on medium and small particle datasets, testing patch sizes of $64^3$ and $16^3$ voxels with corresponding stride adjustments (Table~\ref{tab:explicit_implicit}). For both groups, we initialized training with correctly identified particles from watershed-based segmentations and performed a single SVL iteration. Unlike prior experiments, medium and small particle groups were treated independently, without merging them into a combined training set.

\begin{table*}[!t]
\centering
\caption{\label{tab:explicit_implicit} Quantitative comparison of Explicit and Implicit Boundary Detection}
\setlength{\tabcolsep}{6pt}
\begin{tabular}{ccccccccccc}
\toprule
{} & \multicolumn{3}{c}{Medium} & \multicolumn{3}{c}{Small}\\
\midrule
{} & Patch/Stride & Particles (\#) & Training (min) & Inference (sec) & Particles (\#) & Training (min) & Inference (sec) \\
Explicit & $64^3$/32 & 2921 & 1181 & 549 & 44597 & 12 epochs & 1841 \\
Explicit & $16^3$/16 & 2921 & 1027 & \textbf{239} & 44507 & 14 epochs & \textbf{1030} \\
\midrule
Implicit & $64^3$/32 & 2891 & 255 & 255 & 47026 & 606 & 1176 \\
Implicit & $64^3$/16 & 2980 & 1996 & 254 & 47454 & 9 epochs & 1182 \\
Implicit & $16^3$/16 & 2976 & \textbf{178} & 256 & 47177 & \textbf{460} & 1231 \\
Implicit & $16^3$/8  & \textbf{2993} & 1348 & 256 & \textbf{47501} & 12 epochs & 1217 \\
\bottomrule
\end{tabular}\\
\raggedright
Results from the first iteration of Self-Validated Learning, initialized using watershed-based segmentations for the medium- and small-sized particle groups, respectively. The table compares the performance of implicit and explicit boundary detection across different patch and stride sizes. All models were trained for up to 15 epochs or a maximum of 48 hours (whichever came first) using four NVIDIA V100 GPUs. Bold values indicate the best performance in each column.\\
\par
\end{table*}

\subsubsection{Patch and Stride Settings}
For a given patch size, smaller stride sizes generally improve segmentation quality due to increased overlap between patches during inference. This overlap allows for better aggregation of predictions, averaging probability maps in the explicit case and accumulating boundaries in the implicit case, which helps detect continuous boundaries more reliably. However, this also leads to longer computation times during both training and inference. To balance accuracy and efficiency, we selected stride sizes of $\text{patch size}/2$ for explicit and $\text{patch size}/4$ for implicit boundary detection, ensuring comparable inference runtimes.

As a pragmatic strategy, finer strides could be reserved for later SVL iterations, when only the most difficult particles remain. This approach mirrors our use of full-resolution data in final iterations once most particles have already been segmented.

\subsubsection{Advantages of Implicit Boundary Detection}
Implicit boundary detection significantly outperformed the explicit approach in both accuracy and robustness. This is primarily because implicit training patches inherently contain both foreground and background voxels, mitigating the class imbalance problem. In contrast, explicit methods are challenged by thin and sparse boundary signals and require additional techniques such as class balancing \cite{BUDA2018249}, boundary enlargement (see Eq.~\ref{eq:boundary_enlargement}), or reducing the boundary threshold \cite{wolny_accurate_2020}. These approaches add complexity and require manual tuning, and still fail to fully resolve the imbalance.

Moreover, implicit methods naturally accommodate unlabeled regions, these areas are simply treated as background, whereas explicit methods require an ignore mask, since boundaries in such regions are undefined (Appendix \ref{sec:explicit}).

Interestingly, patch size ($64^3$ or $16^3$ voxels) did not significantly affect performance for the explicit method (Table~\ref{tab:explicit_implicit}). For implicit detection, however, smaller patch sizes yielded better results, mainly because they allowed for finer strides and more localized feature learning.

\subsubsection{Computational Considerations}
The explicit method using a patch size of 16 and a training stride of 16 yielded the shortest inference times. However, this advantage stems from the larger inference stride used (8 for explicit vs. 4 for implicit under the same configuration). When using identical patch and stride settings, implicit training is significantly faster and segmentation performance is better, except in the case with a patch size of 64 and a stride of 32, where the explicit method slightly outperforms the implicit method.

This difference arises because explicit boundary detection requires generating boundary labels for the entire volume and incorporating an ignore mask into the loss function (Appendix~\ref{sec:explicit}). In contrast, implicit training operates only on verified, correctly labeled particles, making it more efficient.

All models were trained for up to 15 epochs or a maximum of 48 hours (whichever came first) using four NVIDIA V100 GPUs. We also adjusted batch sizes based on patch size to optimize GPU memory usage. For a patch size of 64, we retained Biomedisa’s default batch sizes (24 for training and inference). For a patch size of 16, we increased batch sizes to 48 for training and 1024 for inference.

\subsubsection{Additional Findings}
Training solely on the small particle group yielded slightly better performance than training simultaneously on all particle sizes (47,501 vs. 47,228 matched particles; see Table \ref{tab:comparison_methods}). While this difference may result from longer training (12 vs. 8 epochs) or natural statistical variation, it also suggests that medium-sized particles offer little additional benefit for small particle segmentation and that combining both groups creates a more complex optimization challenge.

\begin{figure*}[!t]
\centering
\begin{minipage}{.25\textwidth}
\centering
\includegraphics[width=0.95\textwidth]{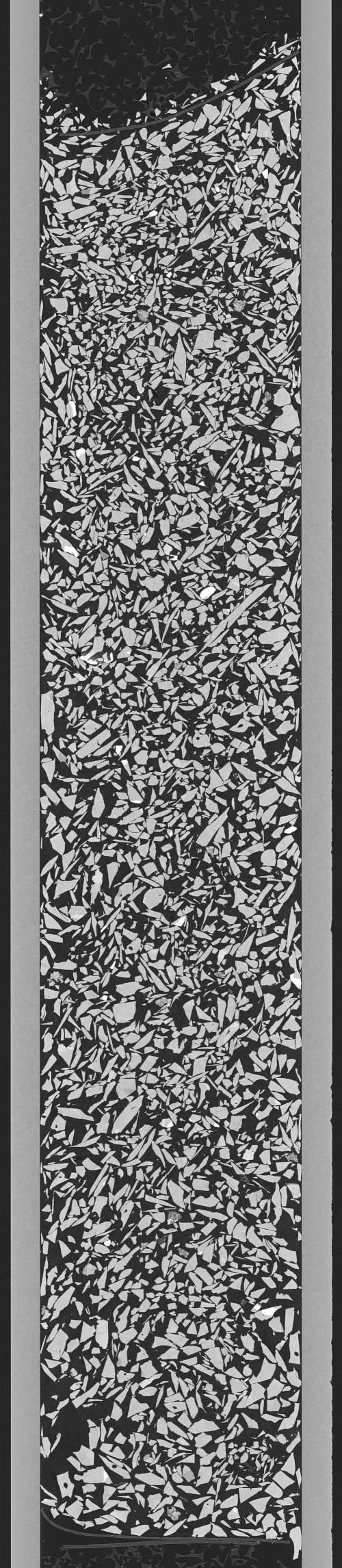}
Tomogram \\ Particle volume: \SI{8.33}{\milli\meter}${}^3$ \\ \phantom{Volume: 12.3\% (4.5\%)}
\end{minipage}%
\begin{minipage}{.25\textwidth}
\centering
\includegraphics[width=0.95\textwidth]{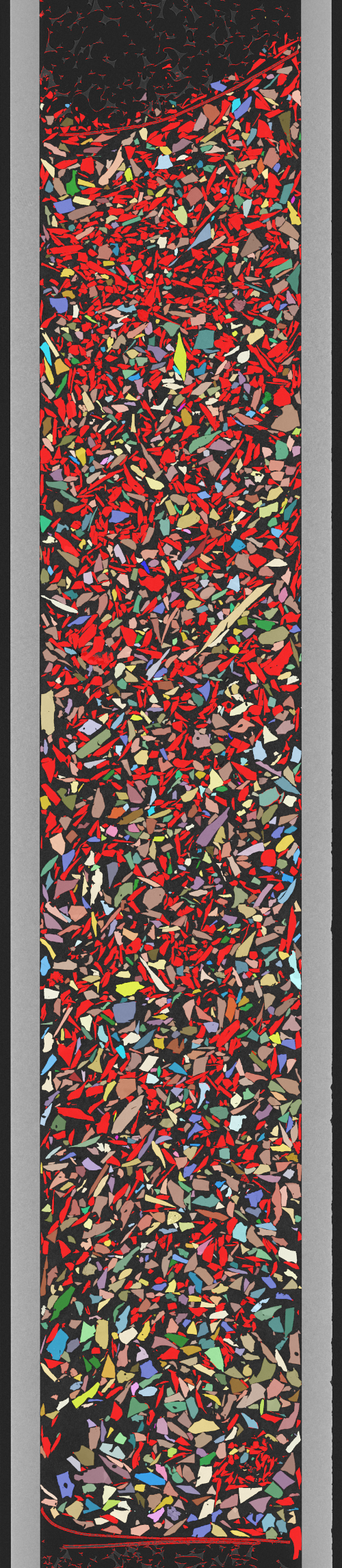}
SAM ViT-L\\ Particles: 23692 (3989) \\ Volume: 66.7\% (7.6\%)
\end{minipage}%
\begin{minipage}{.25\textwidth}
\centering
\includegraphics[width=0.95\textwidth]{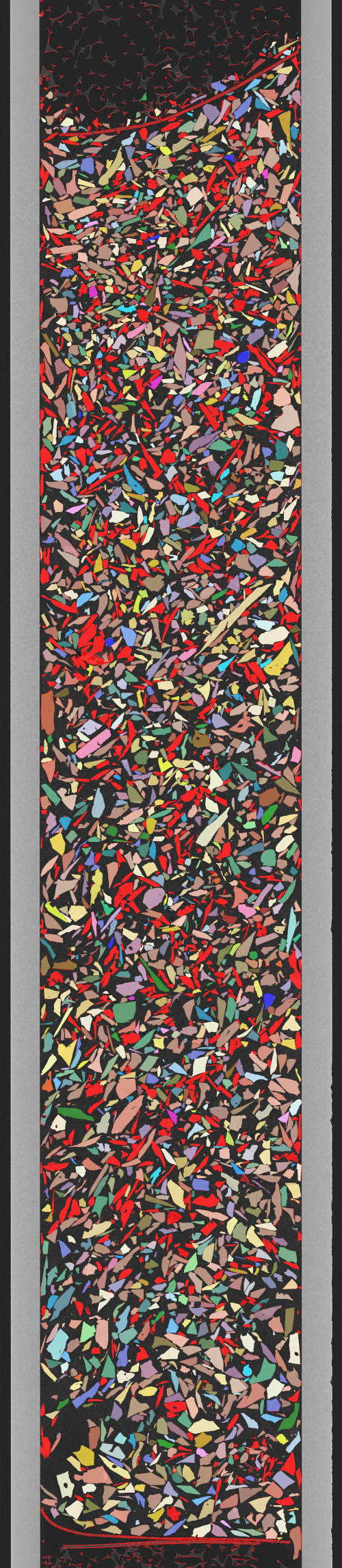}
Advanced Watershed \\ Particles: 36528 (5298) \\ Volume: 83.5\% (5.9\%)
\end{minipage}%
\begin{minipage}{.25\textwidth}
\centering
\includegraphics[width=0.95\textwidth]{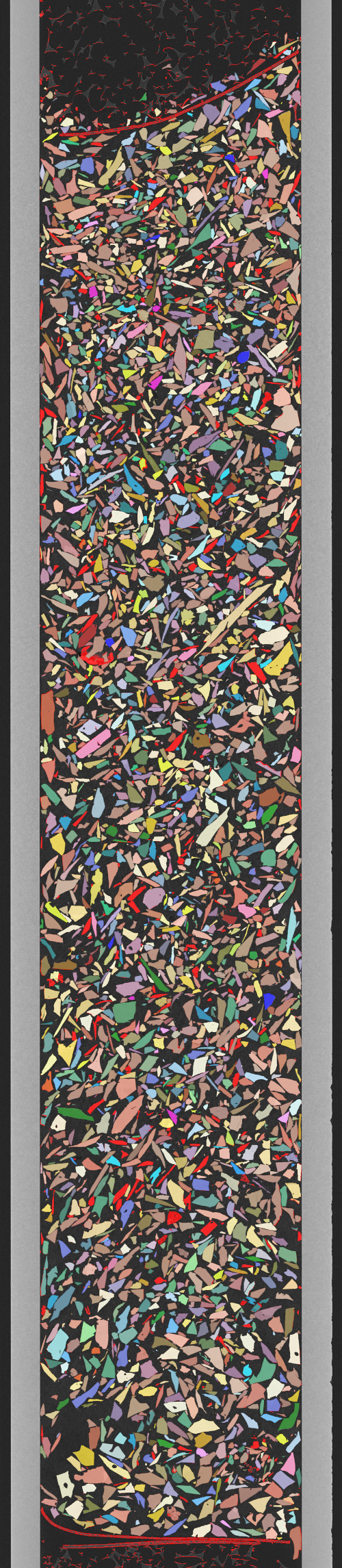}
SVL@2.5 \\ Particles: 48168 (2774) \\ Volume: 95.2\% (1.9\%)
\end{minipage}%
\caption{Visual comparison of different segmentation methods for the class of small particles (between 63 microns and 250 microns). Unmatched particles are shown in red. For each method, the metrics indicate the number of identified particles and the percentage of correctly segmented volume relative to the total particle volume listed in the leftmost column. The first value corresponds to the displayed scan, while the value in parentheses shows additional particles identified exclusively in the other two re-scans of the same sample.}
\label{fig:63microns}
\end{figure*}

\begin{figure*}[!t]
\centering
\begin{minipage}{.48\textwidth}
\centering
\begin{minipage}{.48\textwidth}
\includegraphics[width=\linewidth]{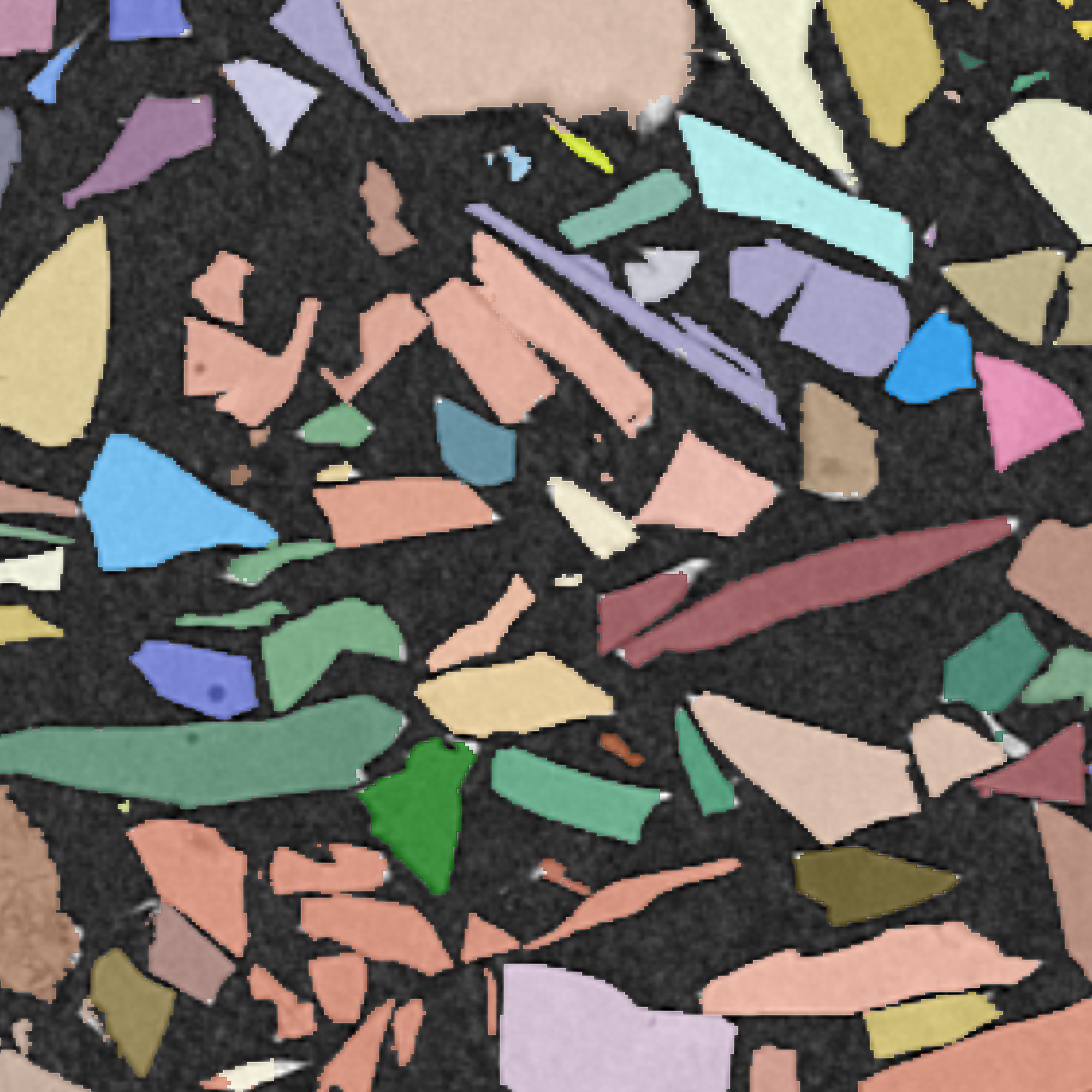}
\end{minipage}%
\hspace{0.02\textwidth} 
\begin{minipage}{.48\textwidth}
\includegraphics[width=\linewidth]{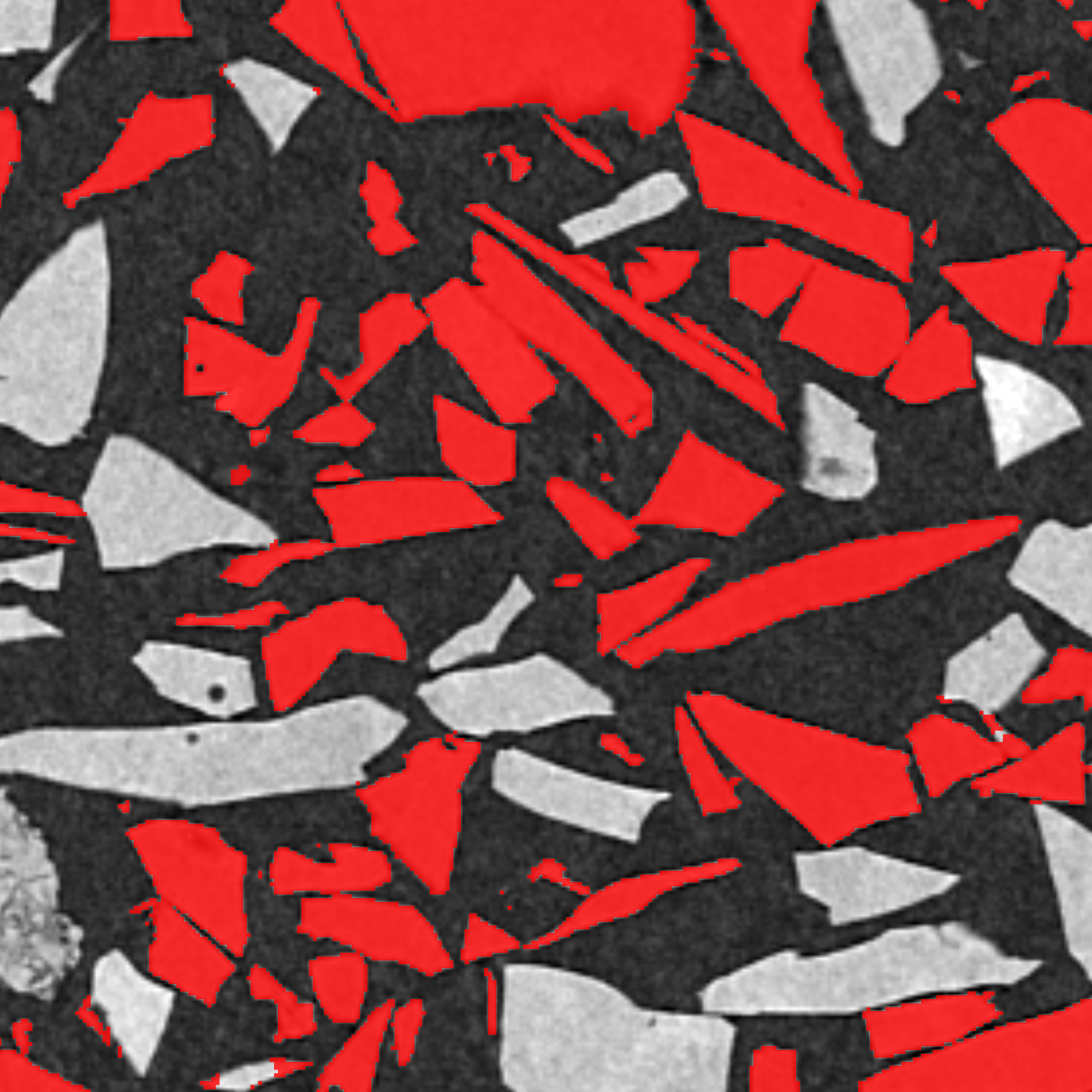}
\end{minipage}
\\ \vspace{0.4em}SAM ViT-L
\end{minipage}%
\hspace{0.01\textwidth} 
\vrule width 0.5pt
\hspace{0.01\textwidth} 
%
\begin{minipage}{.48\textwidth}
\centering
\begin{minipage}{.48\textwidth}
\includegraphics[width=\linewidth]{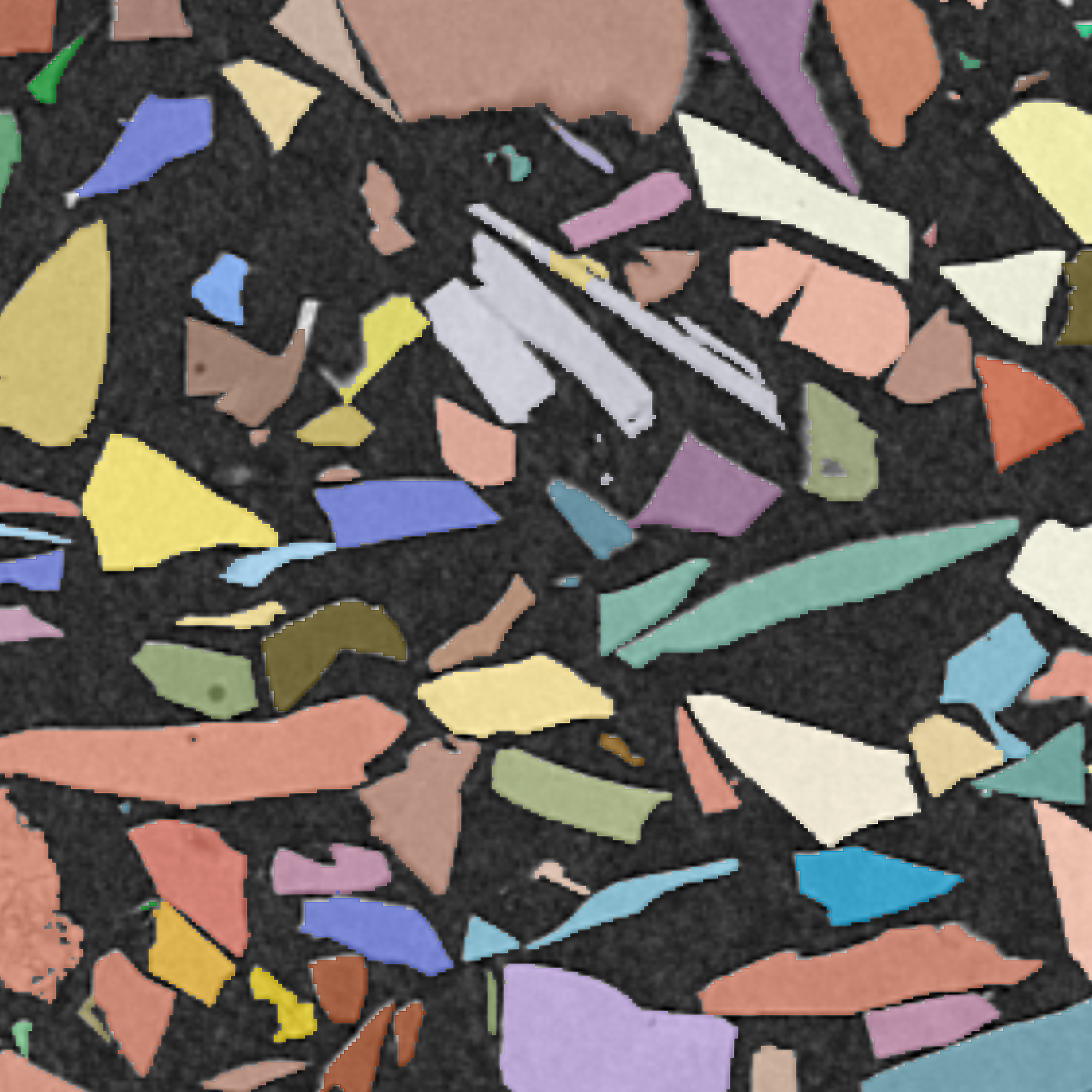}
\end{minipage}%
\hspace{0.02\textwidth}
\begin{minipage}{.48\textwidth}
\includegraphics[width=\linewidth]{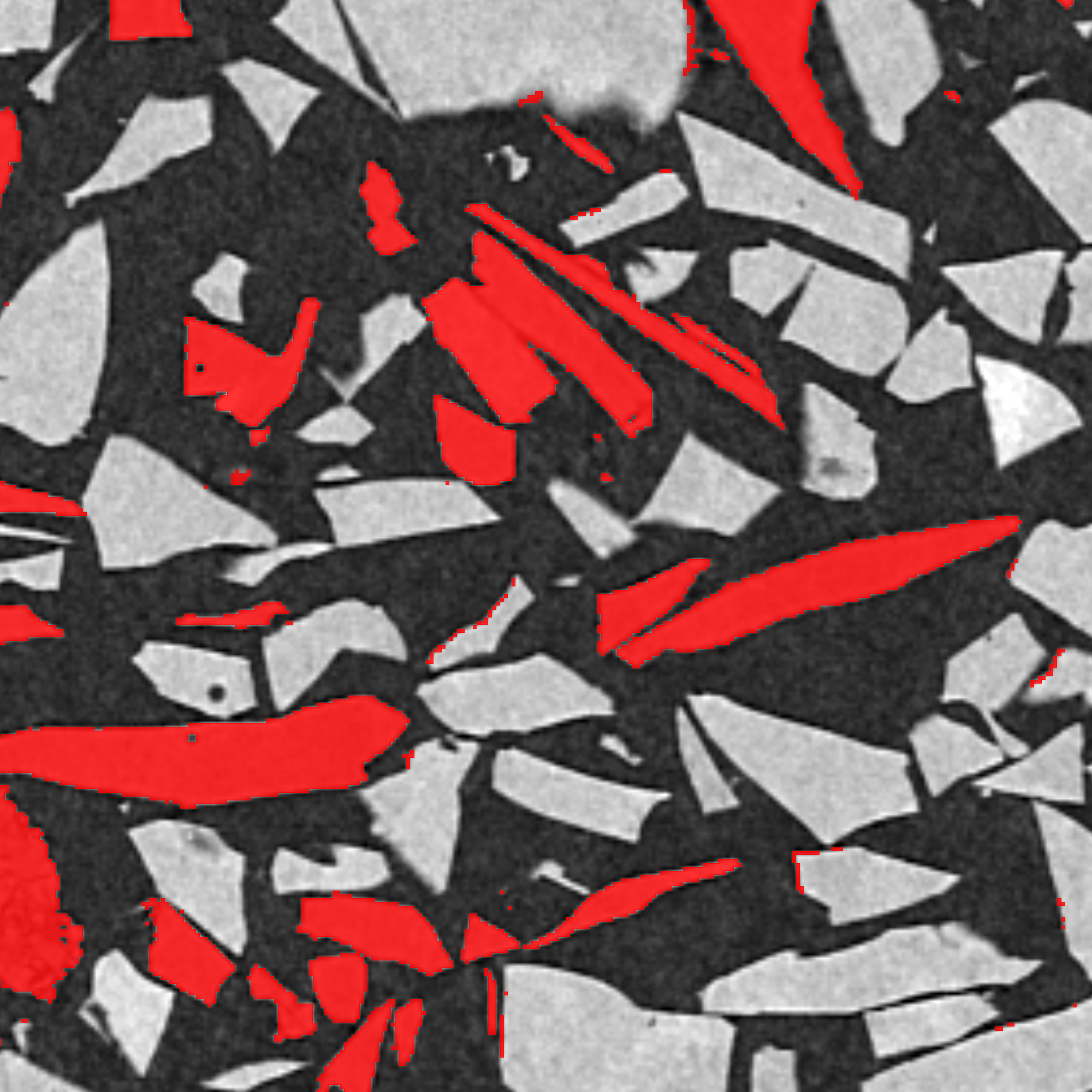}
\end{minipage}
\\ \vspace{0.4em}Advanced Watershed
\end{minipage}
\vspace{1em}

\begin{minipage}{.48\textwidth}
\centering
\begin{minipage}{.48\textwidth}
\includegraphics[width=\linewidth]{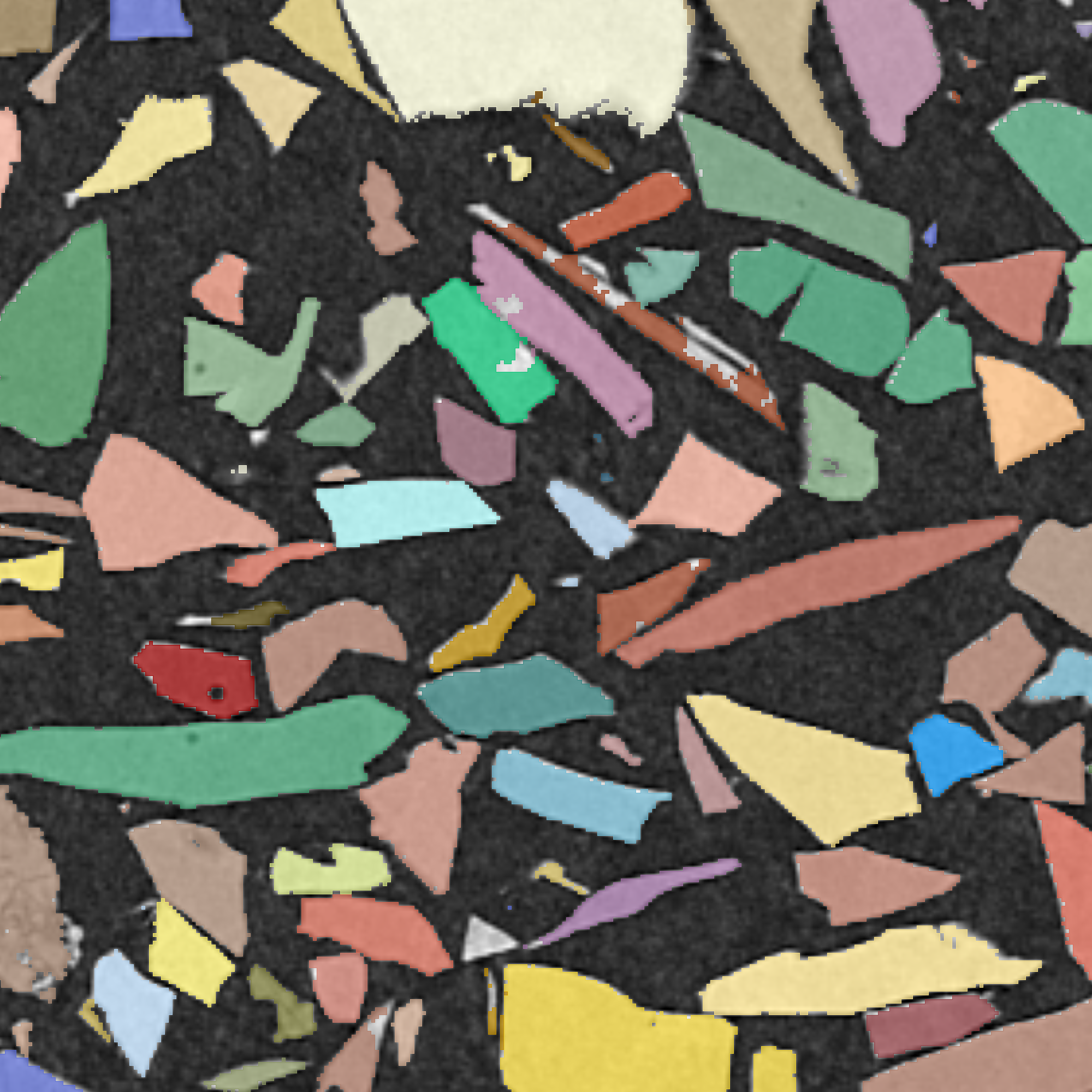}
\end{minipage}%
\hspace{0.02\textwidth}
\begin{minipage}{.48\textwidth}
\includegraphics[width=\linewidth]{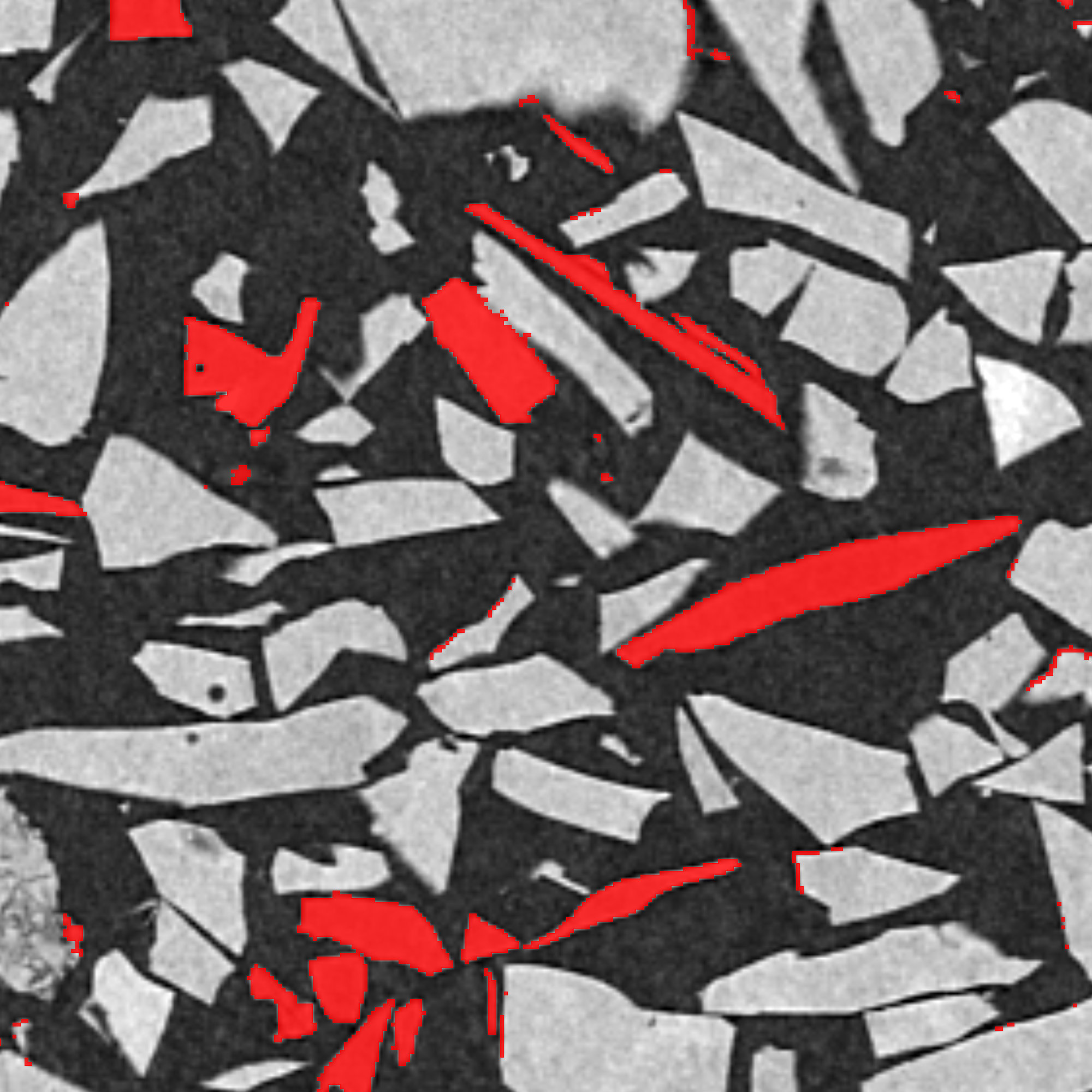}
\end{minipage}
\\ \vspace{0.4em}SVL@1
\end{minipage}%
\hspace{0.01\textwidth}
\vrule width 0.5pt
\hspace{0.01\textwidth}
\begin{minipage}{.48\textwidth}
\centering
\begin{minipage}{.48\textwidth}
\includegraphics[width=\linewidth]{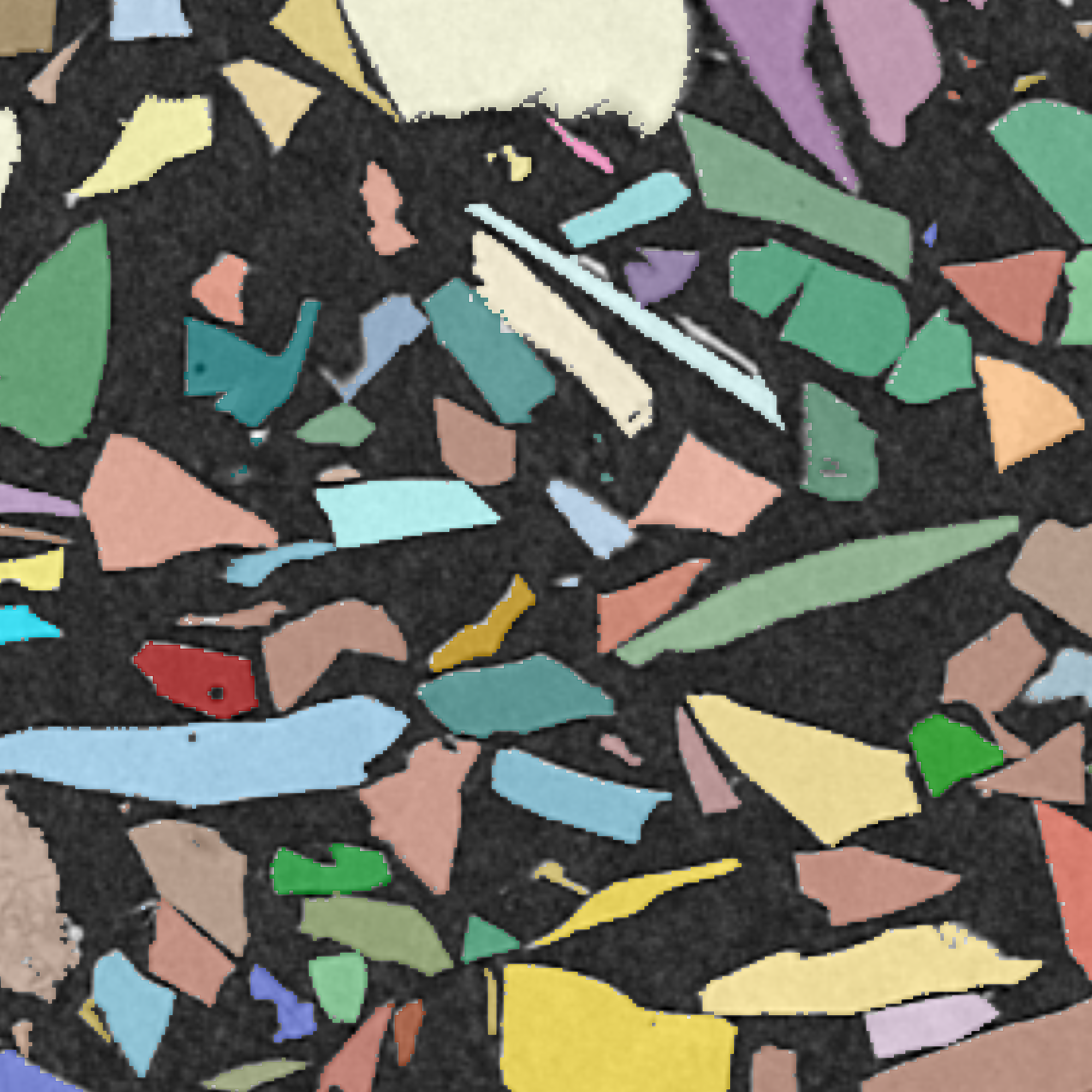}
\end{minipage}%
\hspace{0.02\textwidth}
\begin{minipage}{.48\textwidth}
\includegraphics[width=\linewidth]{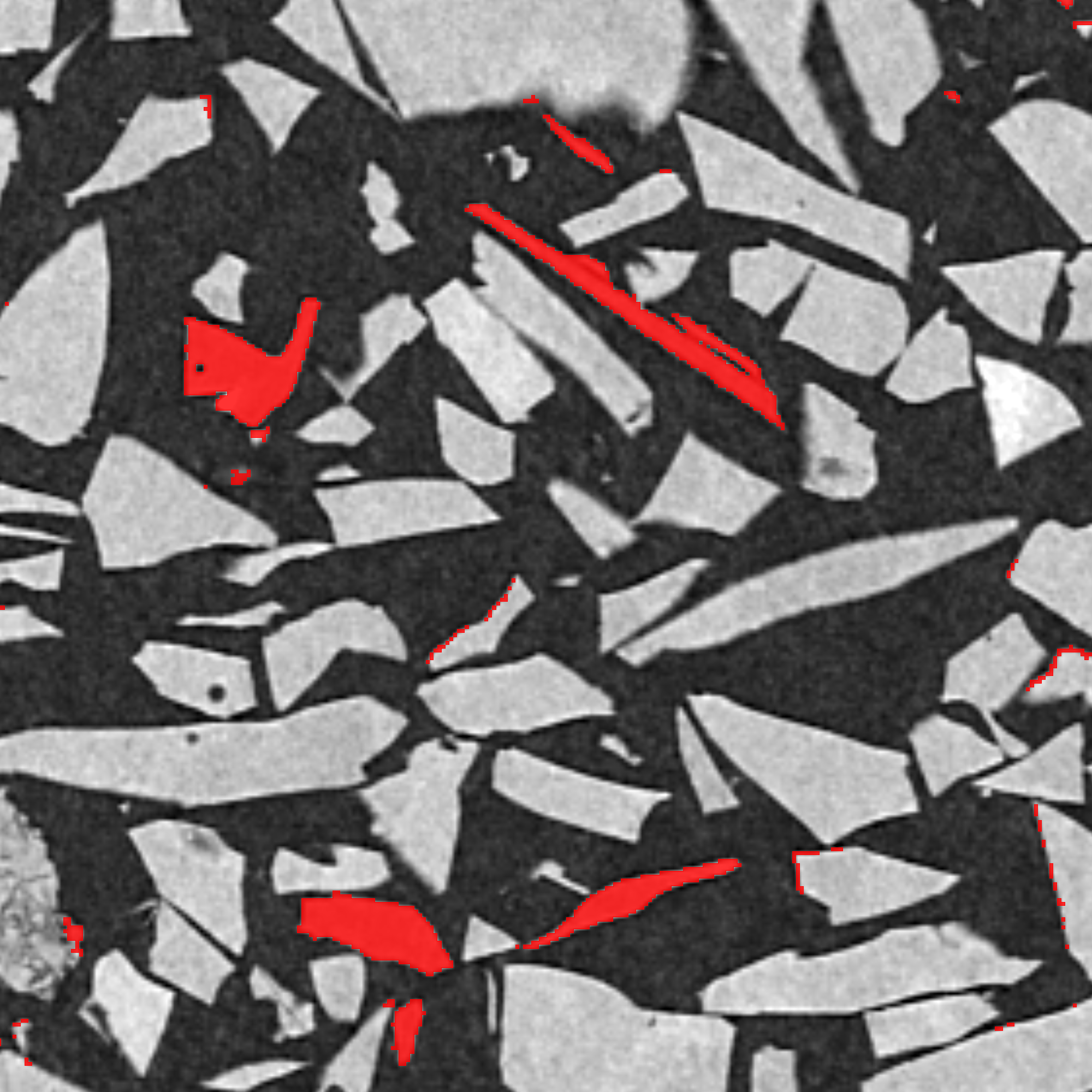}
\end{minipage}
\\ \vspace{0.4em}SVL@2.5
\end{minipage}
\vspace{1em}

\begin{minipage}{.48\textwidth}
\centering
\begin{minipage}{.48\textwidth}
\includegraphics[width=\linewidth]{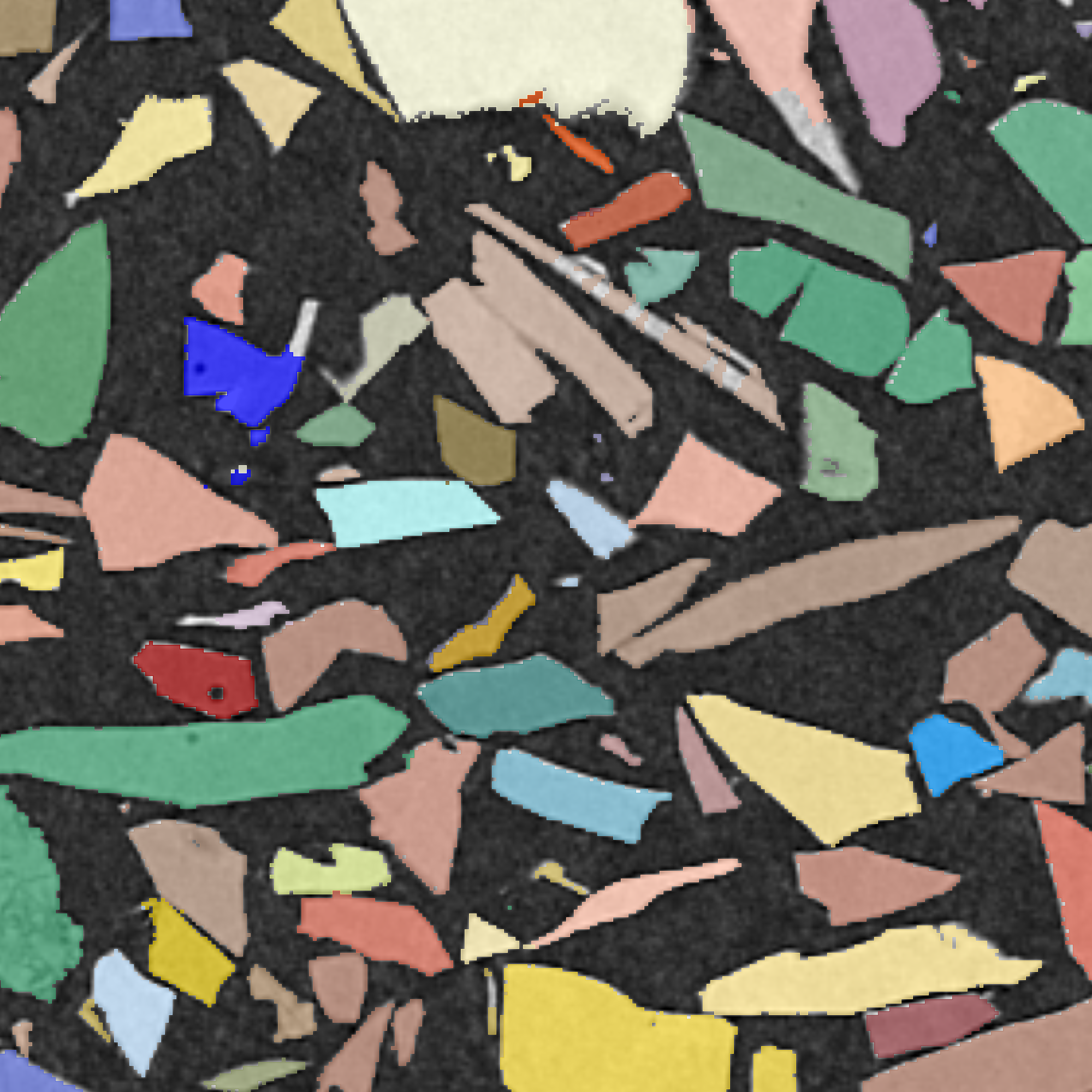}
\end{minipage}%
\hspace{0.02\textwidth}
\begin{minipage}{.48\textwidth}
\includegraphics[width=\linewidth]{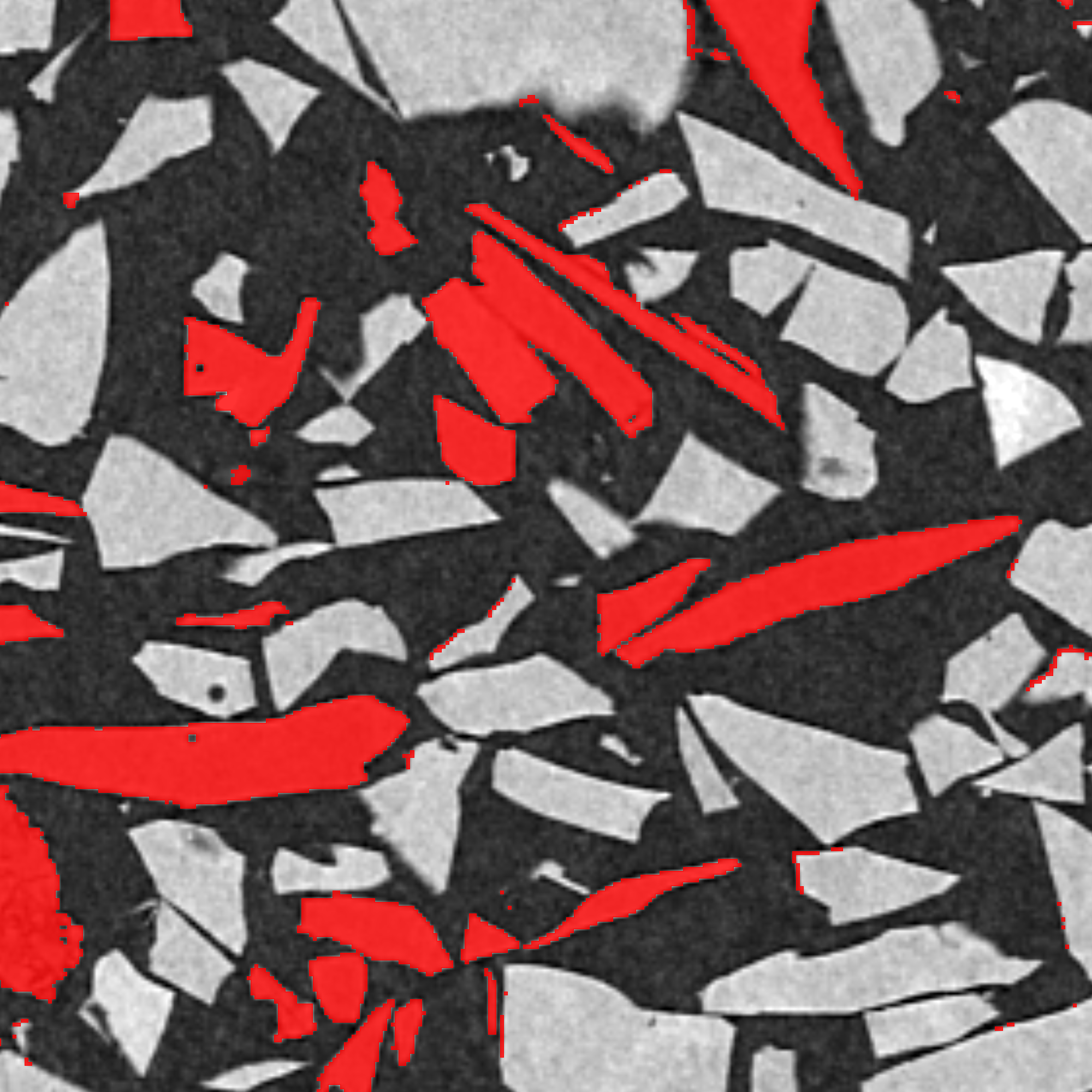}
\end{minipage}
\\ \vspace{0.4em}Explicit@1
\end{minipage}%
\hspace{0.01\textwidth}
\vrule width 0.5pt
\hspace{0.01\textwidth}
\begin{minipage}{.48\textwidth}
\centering
\begin{minipage}{.48\textwidth}
\includegraphics[width=\linewidth]{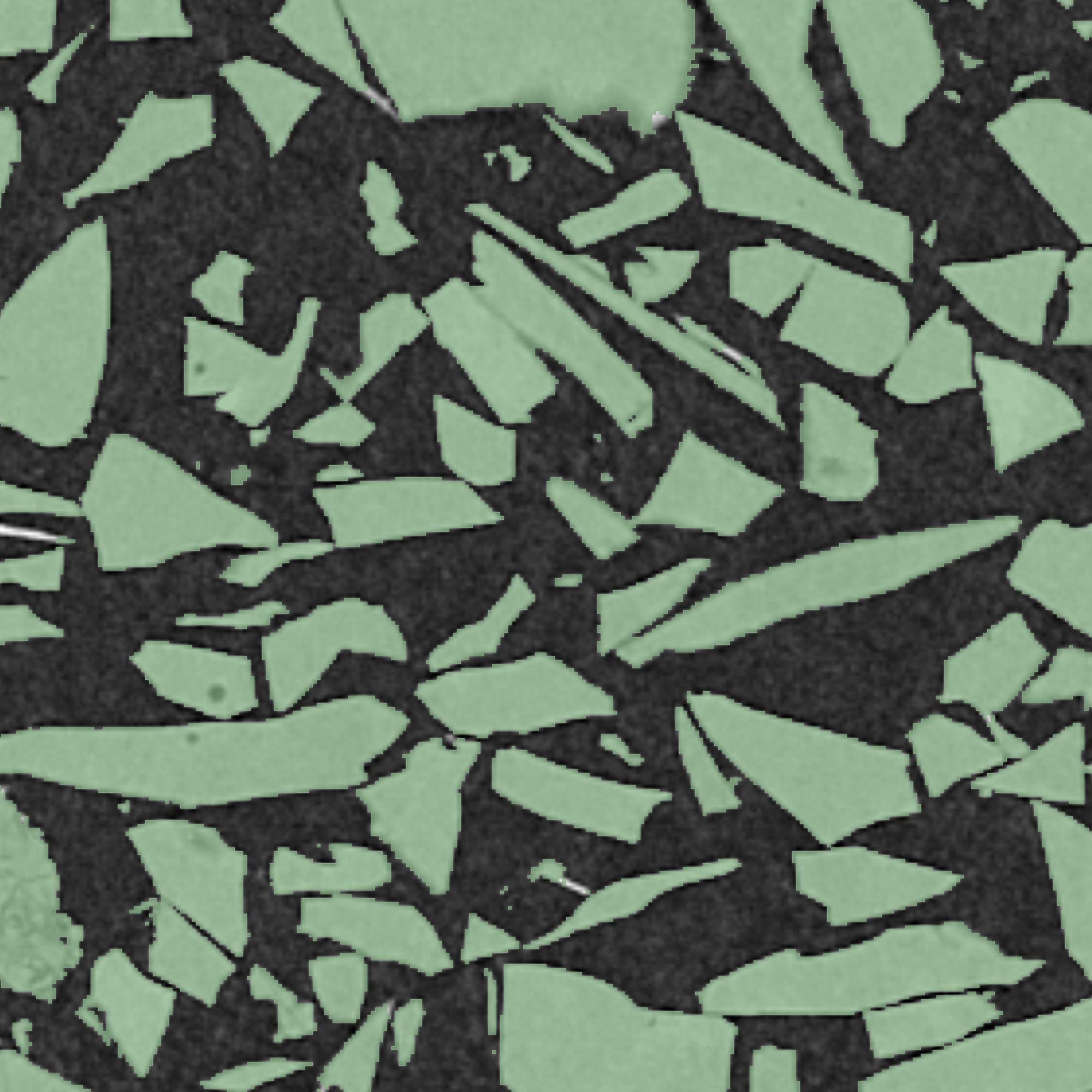}
\end{minipage}%
\hspace{0.02\textwidth}
\begin{minipage}{.48\textwidth}
\includegraphics[width=\linewidth]{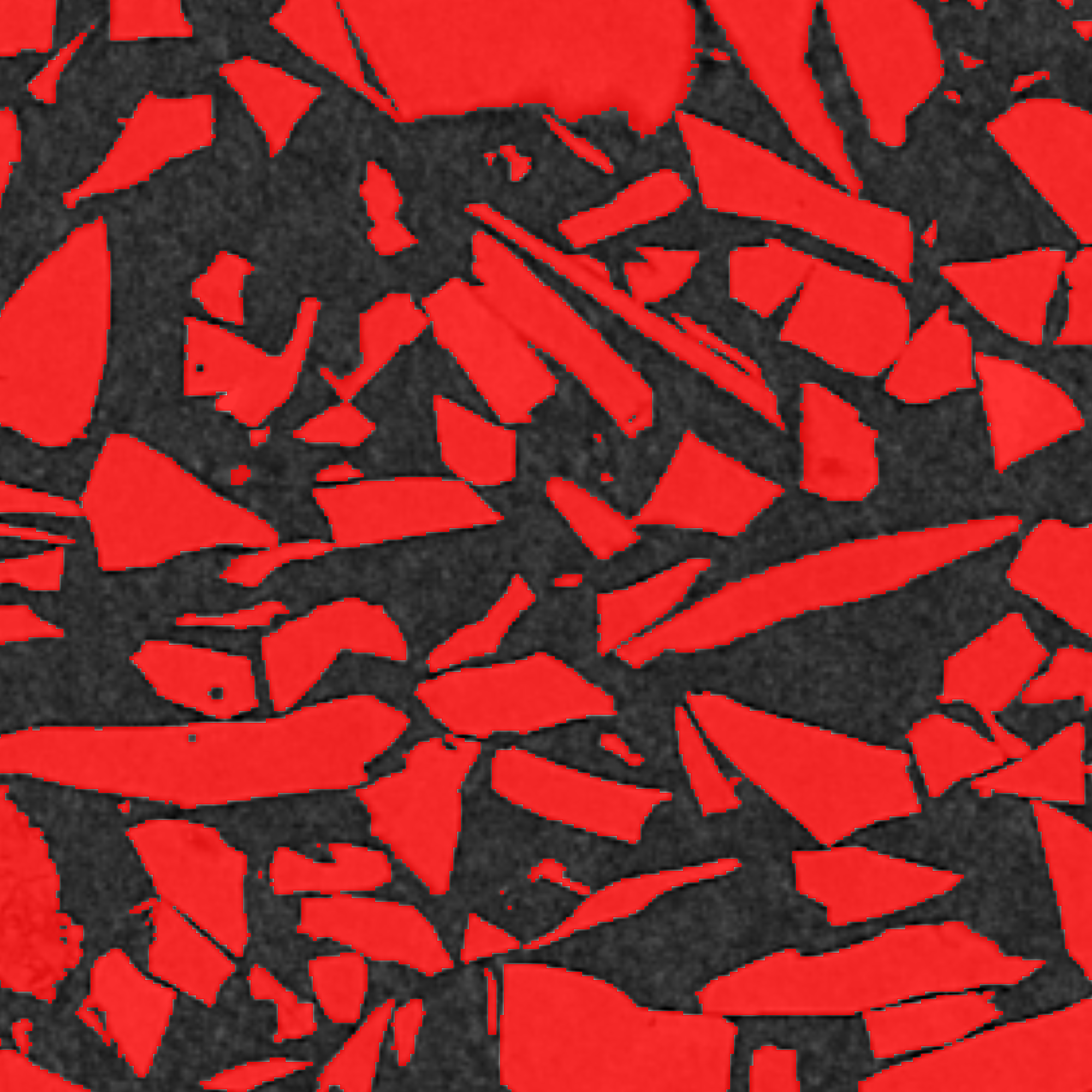}
\end{minipage}
\\ \vspace{0.4em}ST@3
\end{minipage}
\caption{Zoomed-in visual comparison of segmentation results for small-sized particles (\SIrange{63}{250}{\micro\meter}). Each pair shows the initial segmentation before matching (left) and the unmatched particles highlighted in red (right). In the Self-Validated Learning (SVL) framework, boundary detection is applied to the unmatched particles (used as a positive mask) in the subsequent iteration.}
\label{fig:63microns_zoom}
\end{figure*}

\begin{figure*}[!t]
\centering
\begin{minipage}{.25\textwidth}
\centering
\includegraphics[width=0.95\textwidth]{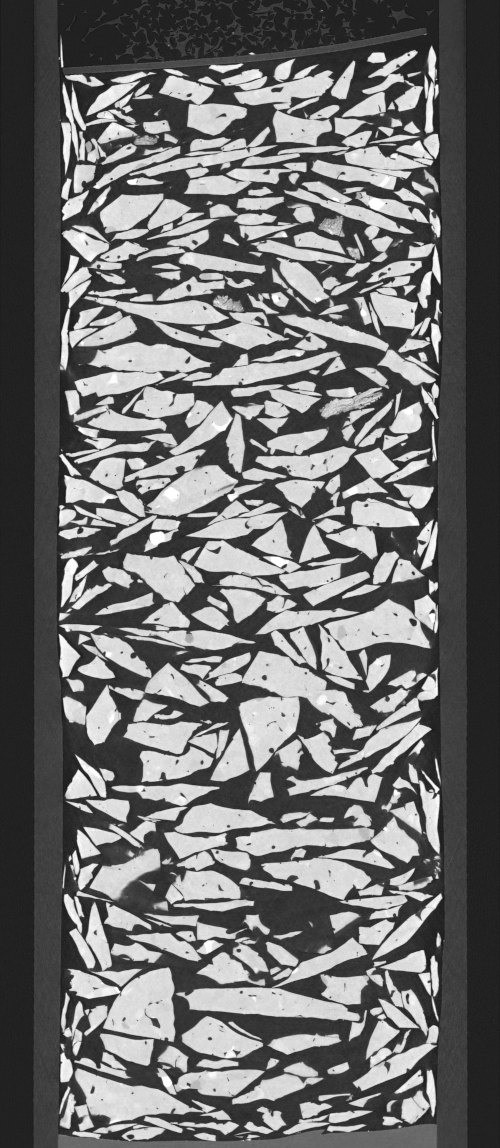}
Tomogram \\ Particle volume: \SI{18.1}{\milli\meter}${}^3$ \\ \phantom{Volume: 12.3\% (4.5\%)}
\end{minipage}%
\begin{minipage}{.25\textwidth}
\centering
\includegraphics[width=0.95\textwidth]{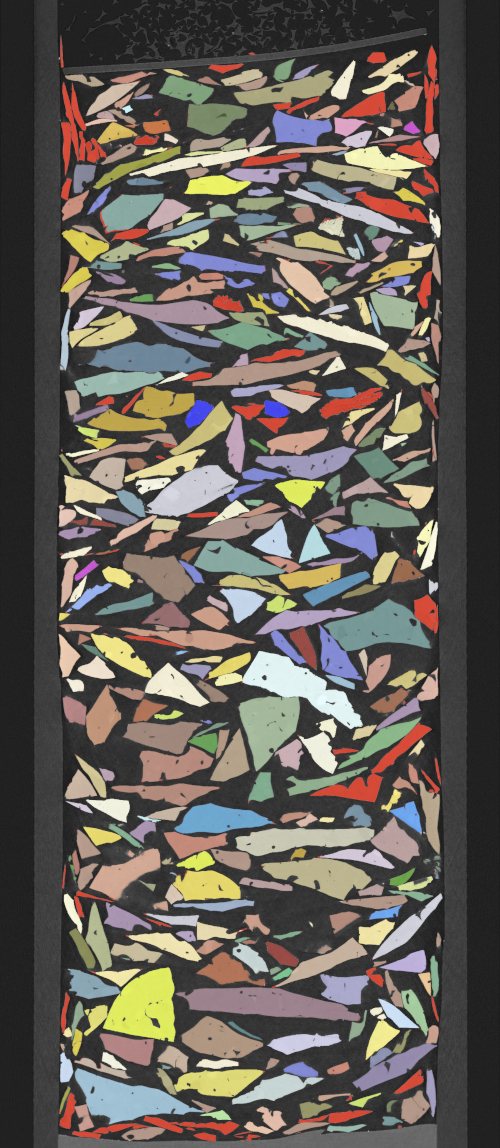}
SAM ViT-L\\ Particles: 2547 (189) \\ Volume: 92.9\% (3.3\%)
\end{minipage}%
\begin{minipage}{.25\textwidth}
\centering
\includegraphics[width=0.95\textwidth]{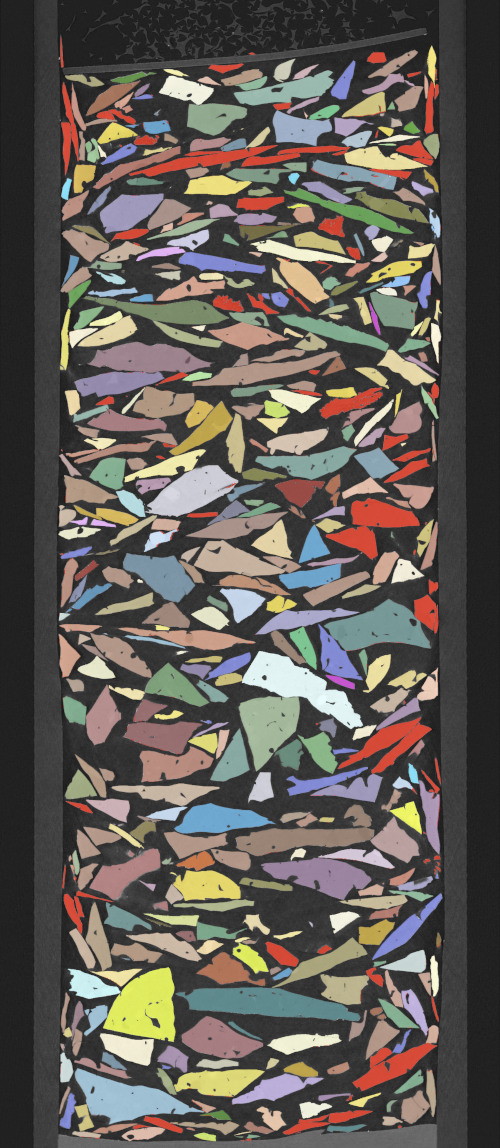}
Advanced Watershed \\ Particles: 2632 (218) \\ Volume: 91.3\% (4.9\%)
\end{minipage}%
\begin{minipage}{.25\textwidth}
\centering
\includegraphics[width=0.95\textwidth]{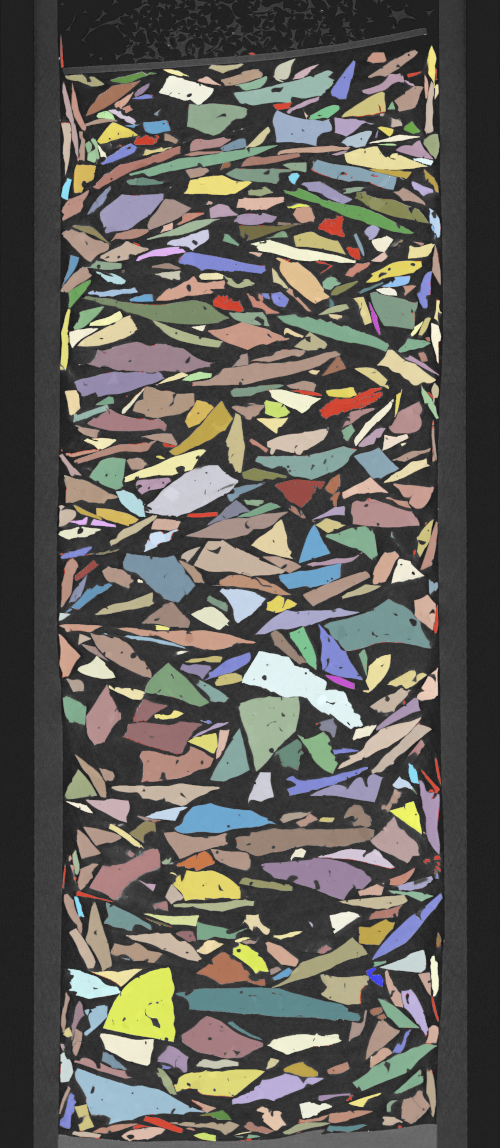}
SVL@2.5 \\ Particles: 2999 (58) \\ Volume: 98.0\% (0.4\%)
\end{minipage}%
\caption{Visual comparison of different segmentation methods for the class of medium-sized particles (between 250 microns and 1 mm). Unmatched particles are shown in red. For each method, the metrics indicate the number of identified particles and the percentage of correctly segmented volume relative to the total particle volume listed in the leftmost column. The first value corresponds to the displayed scan, while the value in parentheses shows additional particles identified exclusively in the other two re-scans of the same sample.}
\label{fig:250microns}
\end{figure*}

\begin{figure*}[!t]
\centering
\begin{minipage}{.25\textwidth}
\centering
\includegraphics[width=0.95\textwidth]{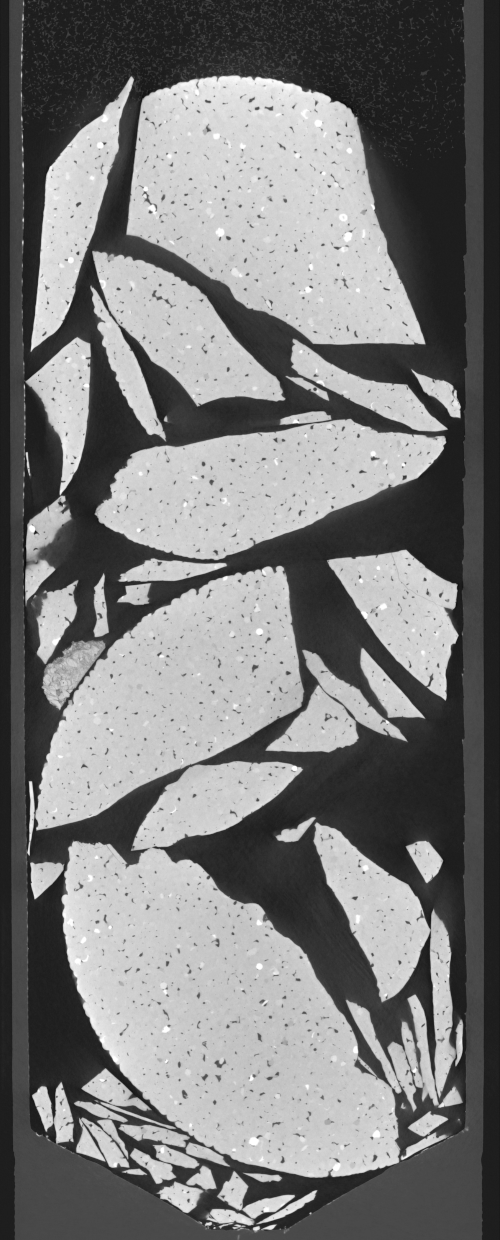}
Tomogram \\ Particle volume: \SI{181.2}{\milli\meter}${}^3$ \\ \phantom{Volume: 12.3\% (4.5\%)}
\end{minipage}%
\begin{minipage}{.25\textwidth}
\centering
\includegraphics[width=0.95\textwidth]{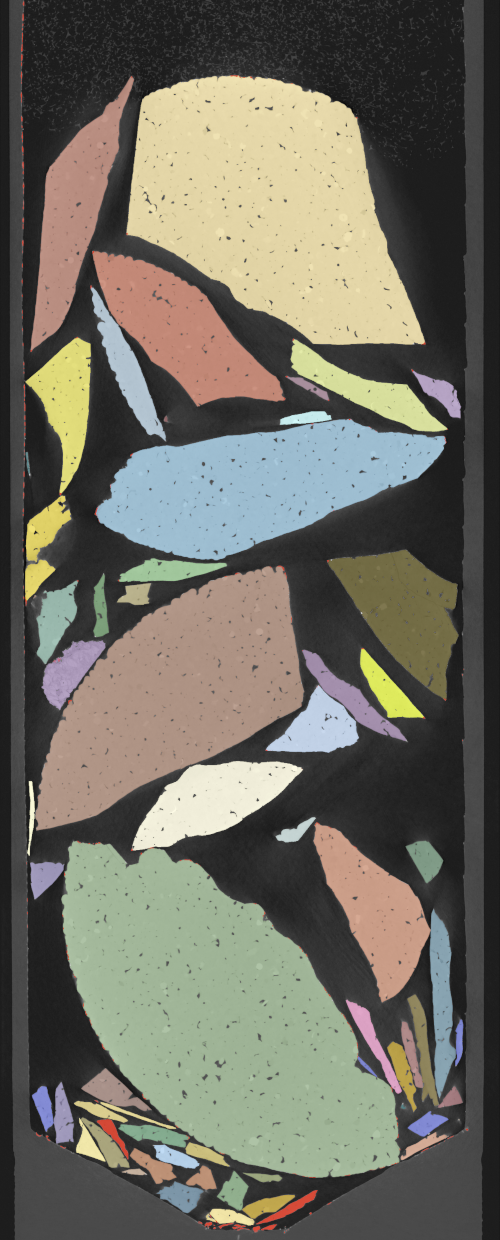}
SAM ViT-L\\ Particles: 188 (4) \\ Volume: 99.5\% (0.1\%)
\end{minipage}%
\begin{minipage}{.25\textwidth}
\centering
\includegraphics[width=0.95\textwidth]{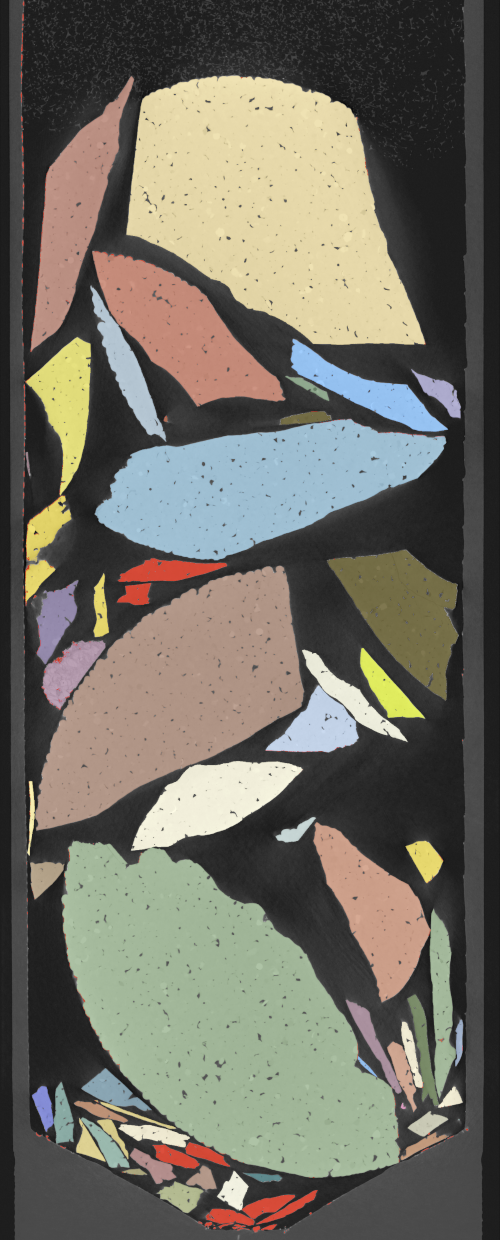}
Advanced Watershed \\ Particles: 168 (11) \\ Volume: 97.9\% (0.8\%)
\end{minipage}%
\begin{minipage}{.25\textwidth}
\centering
\includegraphics[width=0.95\textwidth]{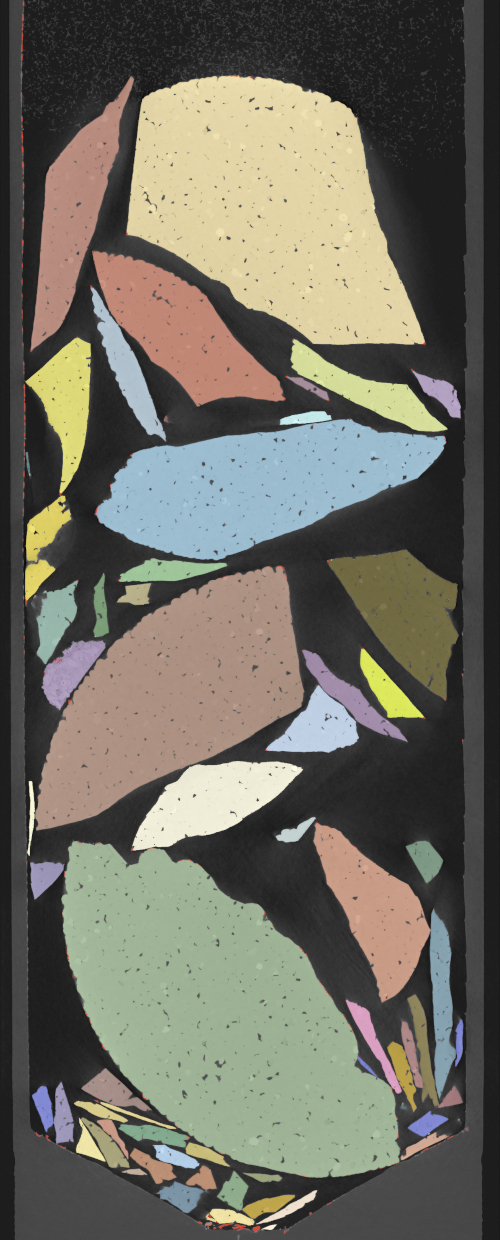}
SVL@2.5 \\ Particles: 226 (4) \\ Volume: 99.7\% (0.0\%)
\end{minipage}%
\caption{Visual comparison of different segmentation methods for the class of large particles (larger than 1 mm). Unmatched particles are shown in red. For each method, the metrics indicate the number of identified particles and the percentage of correctly segmented volume relative to the total particle volume listed in the leftmost column. The first value corresponds to the displayed scan, while the value in parentheses shows additional particles identified exclusively in the other two re-scans of the same sample.}
\label{fig:1mm}
\end{figure*}

\section{Discussion and Future Work}

\subsection{Summary}
In this paper, we extend the concept of self-training from a semi-supervised framework, where human-labeled data is typically essential, to a fully autonomous paradigm. We introduce Self-Validated Learning (SVL), a framework that relies on consistency-based validation of model predictions across reshuffled scans, eliminating the need for manual annotations entirely.

Our method enables the generation of large, diverse, and high-quality datasets of segmented rock particles, paving the way toward a universal model for rock particle segmentation. The resulting dataset offers a strong foundation for developing and evaluating instance segmentation models.

SVL is model-agnostic and applicable beyond our specific U-Net-based implementation. It can also validate predictions from traditional (non-ML) segmentation algorithms or be used to train other architectures, such as SAM, in a similar manner. In cases where even human annotators may struggle to distinguish individual particles, consistency across scans provides strong evidence of correctness, enhancing both label quality and trust.

\subsection{Matching Accuracy and Rotation Dice Score}
A high Rotation Dice score (RotDice) indicates consistent segmentation across reshuffled scans and can therefore serve as a metric for validating segmentation accuracy. In our study, we used a RotDice threshold of 0.9 to accept particle matches, based on empirical observations. Lower thresholds (e.g., 0.85) often led to mismatches across scans, where different particles from one scan were assigned to the same particle in another scan. However, in such cases, one of these mappings typically exhibited a noticeably lower RotDice score. We chose to remove all particles whenever an inconsistent match was detected. Future work might investigate an optimal threshold regarding number of required training iterations by balancing a higher number of accepted particles and their accuracy.

RotDice is also influenced by particle size, as segmentation errors tend to occur at the particle boundaries. Since the surface-to-volume ratio decreases with increasing particle size, larger particles are less affected. To account for this, future work should consider a relative RotDice metric that normalizes for size-dependent effects, or test other metrics such as Average Symmetric Surface Distance (ASSD).

\subsection{Limitations}
Our method relies on generating a positive mask via simple intensity thresholding to focus boundary detection on regions containing particles. In scenarios where such thresholding is not feasible, due to low contrast or heterogeneous backgrounds, a separate model may be required to estimate the foreground mask.

Importantly, a failed match across scans does not necessarily indicate segmentation failure. Thin or ambiguous structures often yield lower Dice scores despite being sufficiently segmented. Future work could incorporate complementary metrics such as Average Symmetric Surface Distance (ASSD) to better assess such cases. Furthermore, a particle might be correctly segmented in only one scan, making consistent matching difficult.

A key limitation of our approach is that systematic over-segmentation errors can persist across all rescans, particularly when internal inclusions are consistently misclassified as separate particles. To mitigate this, we remove objects that are fully enclosed within another particle and reassign their voxels to the surrounding label. However, over-segmentation can also manifest when a particle is consistently split in the same way across scans, for instance, being divided into two identical fragments. In such cases, these fragments are mistakenly treated as correct particles simply because they reappear in each rescan. While such errors could, in principle, be detected by checking whether the affine transformations of the fragments are identical, this step was not performed in the present study.

While increasing the number of reshuffled scans improves the likelihood of capturing and validating all particles, it also raises data acquisition and computational demands, which may pose challenges in resource-constrained environments.

Additionally, our method incurs overhead due to the requirement of identifying and matching particles across scans. Calculating RotDice is computationally intensive and benefits from GPU acceleration. We explored Fast Fourier Transform (FFT)-based methods for estimating rigid body transformations in 2D and limited 3D cases \cite{reddy_fft-based_1996}. Although efficient and robust for specific orientations, these methods do not generalize to full 3D rotations around all three axes.

Finally, while extensive cross-domain validation was beyond the scope of this study, we showed that a model trained on large particles successfully segmented medium and small particles, enabling further training. This promising result suggests that the Self-Validated Learning approach can generalize to other domains without manual input, though this remains to be systematically evaluated.

\subsection{Future Work}
\textit{Incorporating Image Features into Matching:} 
Currently, our particle matching algorithm relies solely on morphological features. For materials exhibiting distinctive intensity patterns, integrating image-based information (e.g., texture or internal contrast) could help improve matching accuracy, particularly in cases where particles have similar shapes.

\textit{Active and Interactive Learning:}
Our approach is well-suited to active learning scenarios. Particles that cannot be confidently matched can be flagged for expert review, enabling targeted manual corrections. This feedback can then be integrated to iteratively refine the model and accelerate convergence, potentially supported by interactive tools such as \textit{nnInteractive} \cite{isensee2025nninteractiveredefining3dpromptable}.

\textit{Toward Generalist and Specialist Models:}
Future work should explore both generalist and specialist directions. On one hand, generalist particle segmentation models trained on large, heterogeneous datasets could enable broad applicability across domains. On the other, specialist models tailored to specific materials (e.g., iron ore, copper ore) may deliver superior performance in constrained settings.

\section{Conclusion}
This study makes three key contributions: (1) the introduction of Self-Validated Learning (SVL) as a fully autonomous self-training framework, (2) the development of a consistency-based validation strategy for evaluating particle segmentations across physically reshuffled scans, and (3) the design of an implicit boundary prediction approach for effective instance segmentation of densely packed particles.

SVL enables models to iteratively generate and validate their own training labels without relying on human annotations or auxiliary pretext tasks. Only predictions that pass a correctness check, based on cross-scan consistency, are retained for further training. This built-in quality control mechanism prevents the accumulation of label noise, a common failure point in conventional pseudo-labeling and self-training approaches.

A major challenge in real-world machine learning is the lack of robust generalization due to limited, biased, or inconsistently labeled datasets. Manual annotation is not only costly and time-consuming, but also prone to subjectivity and inter-observer variability. SVL addresses this by eliminating the need for manual labeling while ensuring label accuracy, offering a scalable and reproducible solution for challenging segmentation problems across domains such as geoscience, materials science, and biomedical imaging.

To maximize performance in particle segmentation, we recommend acquiring at least three scans of the same sample with particles reshuffled between scans. This configuration enables reliable validation through self-matching. While not essential, using a high-contrast container simplifies foreground masking, accelerates processing, and improves segmentation quality.

Importantly, our framework is model-agnostic. It can be applied to traditional algorithms as well as modern deep learning models such as U-Net or SAM, and is readily adaptable to other domains where spatial consistency across observations is available. By enabling the autonomous creation of large, diverse, and validated training datasets, Self-Validated Learning provides a powerful path toward domain-adaptive, annotation-free training in real-world instance segmentation.

\appendices
\section{}

\subsection{Sample Preparation and CT scanning}\label{sec:sampleprep}
A single \SI{8}{\milli\meter} quartz sphere, composed of \SIrange{15}{25}{\micro\meter} quartz beads fused by sintering at 700 degrees for 110 minutes, was mechanically crushed in a drop-weight test following the procedure of Barron et al. \cite{barron_techniques_2025}. The sample was originally prepared to study the relationship between porosity and breakage behavior. The selected sphere had a porosity of less than 5\,\%.

The resulting fragments were brushed into a container and sieved into four size fractions: \textbf{large} (\SI{>1}{\milli\meter}, \cref{fig:matching,fig:1mm}), \textbf{medium} (\SI{1}{\milli\meter}–\SI{250}{\micro\meter}, \cref{fig:segmentation,fig:250microns}), \textbf{small} (\SIrange{250}{63}{\micro\meter}, \cref{fig:63microns}), and \textbf{fines} (\SI{<63}{\micro\meter}, excluded from this study).

Sieving ensures more uniform packing and reduces reconstruction artifacts caused by varying densities when multiple particle sizes are mixed.

The large, medium, and small particle fractions were each packed into \SI{4}{\milli\meter} PEEK tubes and stabilized with foam. Each packed sample was scanned three times, with the particles removed, physically reshuffled, and repacked between scans to enable cross-scan consistency validation. Scanning was conducted at the National Laboratory for X-ray Micro Computed Tomography (CTLab) based at the Australian National University (ANU) using a high-resolution HeliScan system with an optimized space-filling X-ray source trajectory~\cite{Heliscan2018}. The resulting isotropic resolutions were 5.38, 2.8, and \SI{1.54}{\micro\meter}/voxel for the large, medium, and small particle groups, respectively.

\subsection{Segment Anything Model (SAM)}\label{sec:sam}
The Segment Anything Model (SAM) is a vision foundation model designed for interactive 2D image segmentation \cite{kirillov_segment_2023}. To adapt SAM for 3D segmentation, we applied the ViT-L variant using the PyTorch backend \cite{paszke_pytorch_2019} to orthogonal 2D slices along all three axes. The resulting 2D segmentations were used to extract boundary maps, which were subsequently aggregated into a 3D boundary representation. From this point, the same processing pipeline was applied as for boundaries predicted via our implicit boundary detection approach.

To accommodate the large extent along the $z$-axis, we implemented a tiling strategy with a 256-pixel overlap for the $xz$- and $yz$-planes. Additionally, the data was cropped to the foreground particle mask, substantially reducing computation times from 10,272, 10,535, and 40,425 seconds to 10,175, 8,607, and 31,346 seconds for the large, medium, and small particle datasets, respectively, by parallelizing slice processing across four NVIDIA V100 GPUs. This preprocessing step also mitigated downsizing effects caused by SAM's native input resolution of $1024 \times 1024$ pixels and significantly improved the number of matched particles from 192, 2,674, and 13,012 to 192, 2,736, and 27,681, respectively.

Although tiling in the $xy$-plane could potentially further improve segmentation accuracy, we avoided this due to the already high computational cost. While SAM2 \cite{ravi2024sam2segmentimages} is reported to be computationally more efficient, its performance on tomographic images remains mixed: Isensee et al.\ report improved accuracy for interactive segmentation tasks \cite{isensee2025nninteractiveredefining3dpromptable}, whereas Sengupta et al.\ observed reduced segmentation performance on medical CT data \cite{Sengupta_2025}.

\subsection{Advanced Watershed}\label{sec:watershed}
Data was cropped manually. The watershed algorithm was applied to full resolution.

\subsection{Training}\label{sec:training}
Implicit boundary prediction was performed using the 3D U-Net implementation in Biomedisa~\cite{losel_introducing_2020}, built with Keras~\cite{chollet2015keras} and a TensorFlow backend~\cite{abadi_tensorflow_2016}, following the architecture proposed in~\cite{navab_u-net_2015,cicek_3d_2016}. We employed a reduced patch size of $16\times16\times16$ voxels, with stride sizes of 8 for training, 16 for validation, and 4 for inference, which achieved the best performance in our tests (Section \ref{sec:implicit_explicit} and Table \ref{tab:explicit_implicit}).

The overlap between adjacent patches, introduced by using stride sizes smaller than the patch size, enhances segmentation accuracy by allowing structures to be evaluated multiple times. Rather than averaging probability maps, as is commonly done, we aggregate the resulting boundary predictions directly.

For each SVL iteration, we trained for 8 epochs on four NVIDIA V100 using 80\,\% of the volume for training and 20\,\% for validation, with a different validation block selected in each iteration. To accelerate convergence, model weights were initialized from the previous SVL iteration.

Unlike the default Biomedisa configuration, which is tailored for semantic segmentation tasks on downsampled volumes, such as honeybee brains~\cite{losel_natural_2023}, cellular structures~\cite{erozan_automated_2024}, or abdominal structures in medical imaging~\cite{losel_hedi_2023}, our approach operates directly on the original resolution. Instead of extracting patches from resized volumes of $256\times256\times256$ voxels, we enabled the “no scaling” option to preserve full spatial detail. This is essential for accurate instance-level boundary detection, as previously demonstrated for fine root segmentation~\cite{van_der_bom_synchrotron-based_2025} and whole-body reconstructions of ants~\cite{katzke_2025} from large synchrotron micro-CT scans.

Rather than normalizing the entire volume, we applied patch-wise normalization by subtracting the patch mean intensity and dividing by its standard deviation. To improve generalization, we augmented the training data by randomly flipping or swapping all three axes.

\subsection{Multi-GPU Inference}\label{sec:inference}
Boundary prediction was parallelized by distributing consecutive volumetric blocks ($x$-size $\times$ $y$-size $\times$ patch size) across available GPUs using Open MPI. In the $z$-direction, we used the same stride as for patch extraction. Final predictions were computed by averaging overlapping probabilities for explicit boundaries, or by aggregating for implicit boundary detection (Eq.~\ref{eq:aggregating}). This multi-GPU setup scales efficiently with the number of GPUs from both NVIDIA and AMD, as shown in Table~\ref{tab:GPU_comparison}. Most of our experiments were conducted on four NVIDIA V100 GPUs, which are now surpassed by newer GPU generations, such as the NVIDIA GeForce RTX 4080. Leveraging these more advanced architectures significantly reduce training and inference times (Table~\ref{tab:GPU_comparison}).

\begin{table*}[!t]
\centering
\caption{\label{tab:GPU_comparison} Scalability and Inference Time (sec) Across GPU Types and Counts for Implicit Boundary Detection}
\setlength{\tabcolsep}{6pt}
\begin{tabular}{ccccccccccc}
\toprule
GPU(s) & Large & Medium & Small \\
\midrule
4 NVIDIA V100 & 114 & 273 & 1274 \\
4 NVIDIA A4000 & 82 & 201 & 920 \\
4 AMD MI210 & 111 & 231 & 1154 \\
\midrule
1 NVIDIA V100 & 297 & 855 & 4374 \\
1 NVIDIA A4000 & 210 & 664 & 3462 \\
1 NVIDIA 4080 & 116 & 375 & 2080 \\
1 AMD MI210 & 226 & 764 & 4222 \\
\bottomrule
\end{tabular}\\
\raggedright
Inference times for SVL@1 (as used in Section~\ref{sec:svl_results} and Table~\ref{tab:comparison_methods}) across all three particle groups (large, medium, and small) evaluated using different GPU configurations (NVIDIA and AMD, with both single and multi-GPU setups).\\
\par
\end{table*}

\subsection{Separation}\label{sec:separation}
In each SVL iteration, the mask and the previous result are used to identify unmatched particles, i.e., unsegmented regions, which serve as a {\it positive mask} for inference. Only patches containing positive values in this mask are processed. The positive mask is progressively reduced by removing areas of already segmented particles.

Particle separation is performed by removing the predicted boundary from the mask, segmenting connected regions, and assigning each distinct object a unique label. The removed boundary areas are then filled using the nearest label value (via Euclidean distance). Particles are relabeled in ascending order based on size, and small objects are removed according to a user-defined threshold. While not required, this step accelerates the matching process. Finally, the newly segmented objects are added to the previously segmented particles.

\subsection{Explicit Boundary Prediction:}\label{sec:explicit}
Explicit boundary segmentation follows the principles outlined by Wolny et al. \cite{wolny_accurate_2020}, but was implemented in Biomedisa using an ignore mask in the loss function to exclude unmatched regions in each iteration. Unlike the original approach, we did not apply watershed to the boundary probability maps; instead, we directly used the predicted boundaries to segment connected regions into individual particles, as done in our implicit boundary detection method.

Explicit boundary detection trains the 3D U-Net in Biomedisa to predict binary particle boundaries. The input volume is divided into overlapping 3D patches of $64 \times 64 \times 64$ voxels, using a stride of 32 during training and 64 during validation. At inference, patches are processed with 50\,\% overlap (stride 32), and resulting probability maps are averaged to obtain the final boundary predictions.

The model takes an image patch as input and outputs the corresponding precomputed boundary, $b$, of the particle masks (\cref{fig:segmentation}b). The label image, $s$, is transformed into the boundary image, $b$, by:
\begin{align}\label{eq:boundary_enlargement}
  b(s) =
  \begin{cases}
    1, & \text{if }\Phi(s) \ast G_{\sigma}>0.25, \\
    0, & \text{else},
  \end{cases}
\end{align}
where $\Phi$ converts the labeled volume into a two voxels thick boundary image, $G_{\sigma}$ is the isotropic Gaussian kernel, and $\ast$ denotes convolution, effectively enlarging the boundary signal. We set $\sigma = 1.0$ pixel in our experiments.\\
Given the particle binary mask, $m$, and partially correctly segmented particles, $s$, we use a binary ignore mask (\cref{fig:segmentation}c)
\begin{align}
m_I = 1 - \left(b(m)\lor m - b(s)\lor s\right),
\label{eqn:ignore_mask}
\end{align}
to exclude unannotated regions from the loss computation. Hence, not all particles in the dataset need to be segmented for training, allowing an iterative approach where correctly segmented particles are added while flawed ones remain blank at each step.\\
The loss function minimized in each iteration is:
\begin{align*}
\mathcal{L}=m_I\mathcal{L}_{\mathrm{BCE}}(b_P,b_T),
\end{align*}
where $\mathcal{L}_{\mathrm{BCE}}(b_P,b_T)$ is the binary cross-entropy between the predicted particle boundaries $b_P=\mathrm{Unet}(u)$ and the ground truth boundaries $b_T=b(s_t)$ for iteration step $t$, with $m_I$ denoting the ignore mask specified in definition (\ref{eqn:ignore_mask}).

\subsection{Visualization}\label{sec:visualization}
Illustrations were created with 3D Slicer v5.7.0~\cite{jolesz_3d_2014}. Slices were chosen either from axial view (\cref{fig:segmentation}) or coronal view (\cref{fig:63microns,fig:1mm,fig:250microns}). The contrast was adjusted to maximize visual separation from background. Opacity of the labels were set to 0.7, and unmatched areas were color coded red with opacity of 0.8. The values were assigned in ascending order after the particles were sorted with respect to their volume. The U-Net diagram in Figure~\ref{fig:segmentation}, created using Matplotlib v3.8.0 \cite{Hunter:2007}, is based on the schematic illustration by Isensee et al. \cite{isensee2025nninteractiveredefining3dpromptable}.

\subsection{Container removal}\label{sec:container_removal}
For 63-micron scans, the voxel values of the container were too similar to be removed by thresholding. While retaining it in the positive mask is possible, it significantly increases inference time, as all patches containing the container must be processed during inference. To mitigate this, we removed the container by identifying the smallest 10-voxel thick ring in each slice that contains only positive values in the mask.

\subsection{Usage of models in Biomedisa}\label{sec:usage}
Our best-performing model (SVL@2.5) is fully integrated with Biomedisa and its 3D Slicer extension, enabling the “separation” feature for automatic particle segmentation. An optional positive mask can be provided to accelerate processing. Implicit boundary prediction can be trained by simply enabling “separation”, no mask is required. The evaluation scheme, based on multiple reshuffled particle scans, is included in the Biomedisa matching module.

\ifCLASSOPTIONcaptionsoff
  \newpage
\fi

\bibliographystyle{IEEEtran}

\bibliography{MyLibrary.bib, ManualLibrary.bib}

\end{document}